\documentclass[useAMS,usenatbib]{mnras}

\usepackage{hyperref}
\usepackage{graphicx,times,epsfig,rotating}

\usepackage[T1]{fontenc}
\usepackage{ae,aecompl}

\usepackage{graphicx}	
\usepackage{amsmath}	
\usepackage{amssymb}	

\usepackage{color}

\def\mnras{MNRAS}
\def\aj{AJ}
\def\aap{A\&A}
\def\apj{ApJ}
\def\apjl{ApJ}
\def\apjs{ApJS}
\def\araa{ARA\&A}

\def\nat{Nature}
\def\physrep{Phys.\ Rep.}

\newcommand{\msun}{\mbox{M$_{\sun}$ }}
\newcommand{\msunend}{\mbox{M$_{\sun}$}}
\newcommand{\lsun}{\mbox{L$_{\sun}$ }}
\newcommand{\lsunend}{\mbox{L$_{\sun}$}}

\newcommand{\cmthree}{\mbox{cm$^{-3}$}}

\newcommand{\htwo}{\mbox{H$_2$}}

\newcommand{\z}{\mbox{$z$}}

\newcommand{\zsim}{\mbox{$z\sim$ }}

\newcommand{\despotic}{\mbox{\sc despotic}}

\newcommand{\cii}{\mbox{\sc [C~\textsc{ii}]}}

\newcommand{\cp}{\mbox{\sc C$^+$}}
\newcommand{\chx}{\mbox{\sc CH$_{\rm x}$}}
\newcommand{\ohx}{\mbox{\sc OH$_{\rm x}$}}

\newcommand{\hep}{\mbox{He$^+$}}
\newcommand{\hi}{\mbox{H~\textsc{i}}}
\newcommand{\ci}{\mbox{C~\textsc{i}}}
\newcommand{\oi}{\mbox{[O~\textsc{i}}]}
\newcommand{\oiatom}{\mbox{O~\textsc{i}}}

\newcommand{\khl}{\mbox{k$_{\rm H,l}$}}
\newcommand{\khh}{\mbox{k$_{\rm H,h}$}}
\newcommand{\khtwol}{\mbox{k$_{\rm H2,l}$}}
\newcommand{\khtwoh}{\mbox{k$_{\rm H2,h}$}}
\newcommand{\lnte}{\mbox{ln($T_e$)}}

\title[The \cii-FIR deficit in galaxies]{A physical model for the \cii-FIR deficit in luminous galaxies}

\author[Narayanan \& Krumholz]{Desika Narayanan$^{1}$\thanks{E-mail:
    dnarayan@haverford.edu} \& Mark
  R. Krumholz$^{2, 3}$\\$^{1}$Department of Physics and Astronomy,
  Haverford College, 370 Lancaster Ave, Haverford, PA
  19041\\$^{2}$Research School of Astronomy \& Astrophysics,
  Australian National University, Canberra, ACT 2611,
  Australia\\$^{3}$Blaauw Visiting Professor, Kapteyn Astronomical Institute, 
  University of Groningen, The Netherlands}
\begin{document}

\date{Submitted to MNRAS}

\pagerange{\pageref{firstpage}--\pageref{lastpage}} \pubyear{2015}

\maketitle

\label{firstpage}

\begin{abstract}
Observations of ionised carbon at $158$ \micron \ (\cii) from luminous
star-forming galaxies at $z \sim 0$ show that their ratios of \cii
\ to far infrared (FIR) luminosity are systematically lower than those of
more modestly star-forming galaxies. In this paper, we provide a theory
for the origin of this so-called ``\cii \ deficit'' in galaxies. Our model
treats the interstellar medium as a collection of clouds with
radially-stratified chemical and thermal properties, which are
dictated by the clouds' volume and surface densities, as well as
the interstellar radiation and cosmic ray fields to which they
are exposed. \cii\ emission arises from the outer,
H~\textsc{i}-dominated layers of clouds, and from regions
where the hydrogen is H$_2$ but the carbon is
predominantly \cp. In contrast, the most shielded regions of clouds
are dominated by CO, and produce little \cii\ emission. This provides
a natural mechanism to explain the observed \cii-star formation relation:
galaxies' star formation rates are largely driven by the surface densities 
of their clouds. As this rises, so does the fraction of gas in the CO-dominated
phase that produces little \cii~emission. Our model further suggests that the apparent
offset in the \cii-FIR relation for high-$z$ sources compared to those at present
epoch may arise from systematically larger gas masses at early times: a
galaxy with a large gas mass can sustain a high star formation rate even
with a relatively modest surface density, allowing copious \cii\ emission
to coexist with rapid star formation.

\end{abstract}

\section{Introduction}
\label{section:introduction}
The $^2P_{3/2}-^2P_{1/2}$ fine structure transition of singly ionised
carbon\footnote{Throughout this paper we will use \cii \ when referring
  to the observable emission
  line, and \cp \ when discussing ionised carbon within the context of
  chemical networks.} (hereafter, \cii) at $\lambda = 158$
\micron \ is one of the most luminous emission lines in star-forming
galaxies, and a principal coolant of the neutral interstellar medium
\citep[ISM;][]{malhotra97a,luhman98a,nikola98a}.  Indeed, \cii \ can
account for $\sim 0.1-1 \%$ of the far infrared (FIR) luminosity in
galaxies \citep{stacey91a}.  The line is excited mainly via collisions
with electrons, neutral hydrogen (\hi), and molecular hydrogen (\htwo),
with the relatively low critical densities of $\sim 44$, $\sim 3\times10^3$,
and $\sim 6 \times 10^3$ \cmthree \ respectively
\citep{goldsmith12a}\footnote{At kinetic temperatures of $8000, 100$
  and $100 $ K, respectively.}.  This, combined with the relatively low
ionisation potential of $11.3$ eV, means that \cii \ emission can arise
from nearly every phase in the ISM.

For present-epoch galaxies, ground-based observations of \cii \ are
challenging owing to telluric water vapor absorption in the Earth's
atmosphere.  Early work with the Kuiper Airborne Observatory (KAO) and
Infrared Space Observatory (ISO) presented evidence of a relationship
between galaxies' \cii \ luminosities and their global star
formation rates (SFRs).  For example, KAO observations of 14 nearby
galaxies by \citet{stacey91a} revealed ratios of \cii/$^{12}$CO (J=1-0)
emission similar to those found Galactic star-forming regions, providing
an indirect link between the \cii \ line luminosity and SFR (via the
$\Sigma_{\rm SFR}-\Sigma_{\rm CO}$ relation in galaxies).  Later
observations utilising ISO by \citet{leech99a} and \cite{boselli02a}
established bona fide relations between \cii \ and the SFR in $\zsim
0$ systems. More recently, the launch of the Herschel Space
Observatory, combined with other high-resolution ultraviolet and
infrared observations, have established a firm \cii-SFR relation in
nearby galaxies
\citep{delooze11a,sargsyan12a,pineda14a,herreracamus15a}.
Cosmological zoom simulations of galaxy formation by \citet{olsen15a}
have suggested that the majority of the \cii \ emission that drives
this relationship originates in molecular gas or photodissociation
regions (PDRs) in giant clouds, providing a natural explanation for
why \cii~should be correlated with star formation.

This said, even since the early days of ISO observations of galaxies,
it has been clear that the SFR-\cii \ relationship breaks down in the
$\zsim 0$ galaxies with the highest infrared luminosities.  Put
quantitatively, the \cii/far infrared (FIR) luminosity ratio decreases
with increasing infrared luminosity, such that ultraluminous infrared
galaxies (ULIRGs; galaxies with $L_\mathrm{IR} \geq 10^{12} L_\odot$)
emit roughly $\sim 10\%$ of the \cii \ luminosity that would be
expected if they had the same \cii/FIR ratios as galaxies of lower FIR
luminosity \citep{malhotra97a,malhotra01a,luhman98a,luhman03b}.  The
evidence for the so-called ``\cii-FIR deficit'' has grown stronger in
the Herschel era.  \citet{graciacarpio11a} showed that the \cii-FIR
deficit is uncorrelated with galaxies' nuclear activity level, and
that similar deficits with respect to FIR luminosity may exist in
other nebular lines as well.  \citet{diazsantos13a} added
significantly to existing samples via a survey of \cii \ emission from
$\sim 250 \ \zsim 0 $ luminous infrared galaxies (LIRGs; $L_{\rm IR}
\geq 10^{11} L_\odot$), and confirmed these conclusions. Other
evidence for this deficit in local systems has come from
\citet{beirao12a}, \citet{croxall12a} and \citet{farrah13a}.  

At high-redshift, the evidence for a \cii-FIR deficit in galaxies is
more mixed.  While there have been a number of \cii\ detections in
$L_{\rm IR} \geq 10^{12} \lsun$ galaxies at $\zsim 2-6$ (see
\citet*{casey14a} for a recent compendium of these data and review of
high-$z$ detections), and certainly many exhibit depressed \cii/FIR
luminosity ratios, many additionally show elevated \cii/FIR luminosity
ratios compared to local galaxies with a similar infrared luminosity
\citep[e.g.][]{iono06a, stacey10a, wagg12a, swinbank12a,
  rawle13a,riechers13a, wang13a,rigopoulou14a,brisbin15a}.

In this paper, we aim to provide a physical explanation for the origin
of the \cii-FIR deficit in heavily star-forming galaxies in the local
Universe, and the more complex pattern found at high redshift.
We do this by developing analytic models for the structure
of giant clouds in galaxies.  We combine chemical equilibrium networks
and numerical radiative transfer models with these cloud models to
develop a picture for how \cii \ emission varies both as a function of
cloud radius, as well as with galactic environment.

Our central argument is relatively straight forward.  Consider a
galaxy with a two-phase neutral interstellar medium comprised of \htwo
\ and \hi. As the surface density of the gas in the galaxy grows,
its star formation rate rises. However, the increased surface density
also increases the ability of the hydrogen to shield itself from dissociating
Lyman-Werner band photons \citep{krumholz08a,krumholz09a,mckee10a},
causing the \htwo/\hi \  ratio to rise.  Within clouds, owing to cosmic
ray and ultraviolet radiation-driven chemistry effects, \cp \ is
prevalent in the photodissociation region but is significantly depleted in
the \htwo \ core.  As a result, the typical decreasing sizes of PDRs
in galaxies of increasing star formation rate results in
proportionally lower \cii \ luminosities.

In what follows, we present a numerical model that shows these
physical and chemical trends explicitly.  In \autoref{section:model}
we describe the model, while in \autoref{section:results} we outline
the main results, including the luminosities of \cii \ in molecular
and atomic gas. In \autoref{section:discussion} we discuss some
applications of this model, including its utility for high-redshift
galaxies and ISM calorimetry.  We additionally discuss the
relationship of our model to other theoretical models in this area, as
well as uncertainties in our model.  Finally, we summarise in
\autoref{section:summary}.

\section{Model Description}
\label{section:model}

Our goal is to explain the observed relationship between \cii~158
$\mu$m emission and star formation rate, for which FIR emission is a
proxy. However, the physical state of a galaxy's ISM obviously depends
on more than its star formation rate. Quantities such as the volume
density, chemical state, and temperature play a role as
well. We therefore develop a minimal model for a galactic ISM as a
whole, and then use that model to compute both SFR and
\cii~emission.

We idealise a galaxy as a collection of spherical, virialised
star-forming clouds. Each cloud consists of several radial zones that each
have distinct column densities, and that are chemically and thermally independent from
one another.  To calculate the line emission from a galaxy, our first
step is to compute the density, column density, and velocity
dispersion of each of these clouds. We do so following the
procedure outlined in \autoref{ssec:phys_struct}. We then use the code
Derive the Energetics and SPectra of Optically Thick Interstellar
Clouds \citep[\despotic;][]{krumholz13a, krumholz13c} to compute the
chemical state (\autoref{section:chemistry}), temperature
(\autoref{ssec:thermal_struct}), and level populations
(\autoref{section:radiative_transfer}) in every layer of the
cloud. The entire model is iterated to convergence as outlined in
\autoref{ssec:convergence}, and, once convergence is reached, we can
compute the total \cii~158 $\mu$m luminosity.

For convenience, we have collected various
parameters that appear in our model in \autoref{table:free_parameters}.

\begin{table*}
\centering
\caption{Parameters used in cloud models
\label{table:free_parameters}
}
\begin{tabular}{@{}lll@{}}
\hline Variable  & Definition & Value \\
\hline
\hline
\multicolumn{3}{@{}l@{}}{Parameters for Cloud Physical Properties (\autoref{ssec:phys_struct})} \\ \hline
$\epsilon_{\rm ff}$ & Dimensionless star formation efficiency & 0.01 \\
$\alpha_{\rm vir}$ & Cloud virial ratio & 1.0 \\
$\rho_{\rm MW}$ & Molecular cloud density normalisation & $2.34\times 10^{-22}$ g cm$^{-3}$ \\
$N$  & Star formation law index & 2  \\
$\phi_{\rm mol}$ & Ratio of molecular to atomic density & $10$ \\
$M_{\rm gal}$ & Galaxy gas mass & $1\times10^9-2\times10^{11}$ \msun \\
$\Sigma_g$ & Cloud surface density & $\sim 50 - 5000$ $M_\odot$ pc$^{-2}$ \\
&&\\ \hline
\multicolumn{3}{@{}l@{}}{Parameters for Cloud Thermal and Chemical Properties} \\ \hline
$N_{\rm zones}$ & Number of radial zones in model clouds & 16 \\
$\chi_{\rm FUV}$ & FUV ISRF & $1.0 \times \mbox{SFR}/(M_\odot\,\mbox{yr}^{-1})$ \\
$\zeta_{-16}$ & cosmic ray ionisation rate & $0.1 \times \mbox{SFR}/(M_\odot\,\mbox{yr}^{-1})$\\
$\alpha_{\rm GD}$ & gas-dust coupling coefficient & $3.2\times10^{-34}$ erg cm$^{3}$ K$^{-3/2}$\\
$\sigma_{\rm d,10}$ & dust cross-section to $10$ K thermal radiation & $2.0 \times 10^{-26}$ cm$^2$ H$^{-1}$\\
$T_{\rm CMB}$ & CMB temperature & 2.73 K \\
$Z'_{\rm d}$ & dust abundance relative to solar & 1 \\
$\beta_{\rm d}$ &dust opacity versus frequency index & 2\\
$A_V/N_H$ & visual extinction per column & $4 \times 10^{-22} Z'$ mag cm$^2$ \\
OPR & ortho to para ratio in \htwo \ gas & 0.25\\
\hline
\end{tabular}
\end{table*}

\subsection{Cloud Physical Structure}
\label{ssec:phys_struct}

The chemical and thermal states of clouds, both in our model and in
reality, will depend upon their volume and column densities, as well
as their velocity dispersions. The first step in our calculation is
therefore to model the relationship between these quantities and
galaxies' star formation rates. To this end, let $\Sigma_g$ be the
surface density of an idealised spherical cloud.  The inner part of
this cloud will be H$_2$-dominated and the outer layers, which are
exposed to the unattenuated interstellar radiation field, will be
dominated by H~\textsc{i}; the specified surface density $\Sigma_g$
includes both of these zones.  The H$_2$-dominated region comprises a
fraction $f_{\mathrm{H}_2}$ of the total cloud mass.
\citet{krumholz08a,krumholz09a,krumholz09b} and \citet{mckee10a}
(hereafter collectively referred to as KMT) show that the molecular
mass fraction for such a cloud obeys
\begin{equation}
\label{equation:kmt}
f_{\rm H_2} \approx 1 - \frac{3}{4}\frac{s}{1+0.25s}
\end{equation}
for $s<2$ and $f_{\rm H_2} = 0$ for $s\geq 2$.  Here $s = \ln
(1+0.6\chi + 0.01\chi^2)/(0.6\tau_{\rm c})$, where $\chi =
0.76(1+3.1Z'^{0.365})$, the dust optical depth of the cloud at
frequencies in the Lyman-Werner band is $\tau_{\rm c} =
0.066\Sigma_g/(\msun {\rm pc^{-2}})\times Z'$, and $Z'$ is the
metallicity normalised to the solar metallicity.  We assume $Z' = 1$
for all model clouds, and thus $f_{\mathrm{H}_2}$ is a function of
$\Sigma_g$ alone.

We relate the atomic and molecular regions via their density
contrast. Specifically, following KMT we define
\begin{equation}
\phi_{\mathrm{mol}} = \frac{\rho_{\mathrm{H}_2}}{\rho_{\mathrm{HI}}},
\end{equation}
where $\rho_{\mathrm{H}_2}$ and $\rho_{\mathrm{HI}}$ are the densities
in the molecular atomic zones, respectively, and we adopt a fiducial value
$\phi_{\mathrm{mol}} = 10$ based on observations. With this choice and
a bit of algebra, one can show that the total cloud mass and radius can
be expressed in terms of $\Sigma_g$ and $\rho_{\rm H_2}$ as
\begin{eqnarray}
\label{equation:mass}
M_c  & = & \frac{9}{16}\pi \left[f_{\rm H_2} + \phi_{\rm mol}(1-f_{\rm H_2})\right]^2 
\frac{\Sigma_g^3}{\rho_{\rm H_2}^2} \\
R_c & = & \frac{3}{4}\left(1 + \phi_{\rm mol}\frac{1-f_{\rm H_2}}{f_{\rm H_2}}\right)
\frac{\Sigma_g}{\rho_{\rm H_2}}.
\end{eqnarray}
We can also express the velocity dispersion of the cloud in terms of these
two variables via the virial theorem. Specifically, we have \citep{bertoldi92a}
\begin{equation}
\alpha_{\rm vir} = \frac{5\sigma_c^2 R_c}{G M_c}
\end{equation}
where $\sigma_c$ is the velocity dispersion and $\alpha_{\rm vir}$ is the virial
ratio. Thus
\begin{equation}
\label{eq:sigmac}
\sigma_c = \sqrt{\frac{3\pi}{20}\alpha_{\rm vir} f_{\rm H_2}^2
\left(1 + \phi_{\rm mol}\frac{1-f_{\rm H_2}}{f_{\rm H_2}}\right) 
\frac{G\Sigma_g^2}{\rho_{\rm H_2}}}
\end{equation}
We adopt a fiducial value for the virial ratio $\alpha_{\rm vir} = 1$, typical of
observed molecular clouds \citep[e.g.][]{dobbs11a,dobbs13a,heyer15a}.
Recalling that $f_{\rm H_2}$ is a function of $\Sigma_g$ alone, we have 
now succeeded in writing the cloud mass, radius, and velocity dispersion in
terms of $\Sigma_g$ and $\rho_{\rm H_2}$ alone, and we have therefore
reduced our model to a two-parameter family.

To proceed further, we now bring star formation into the picture. Consider a
galaxy with a total ISM mass $M_{\rm gal} = N_c M_c$, where $N_c$ is the
number of star-forming clouds in the galaxy. At all but the lowest metallicities, 
stars form only in the molecular region of the ISM 
\citep[e.g.,][]{krumholz11a,glover11a,krumholz12c}. Thus
the total star formation rate of the galaxy is given by
\begin{equation}
\label{eq:sfr}
\mathrm{SFR} = \epsilon_{\rm ff} \frac{f_{\rm H_2} M_{\rm gal}}{t_{\rm ff}},
\end{equation}
where $t_{\rm ff}$ is the free-fall time, and given by
\begin{equation}
\label{eq:tff}
t_{\rm ff} = \sqrt{\frac{3\pi}{32 G \rho_{\rm H_2}}},
\end{equation}
and the quantity $\epsilon_{\rm ff}$ is the fraction of the molecular mass
converted to stars per free-fall time. Observations strongly constrain this
to be within a factor of a few of 1\% \citep{krumholz07b, krumholz12a, 
krumholz14a}, so we adopt $\epsilon_{\rm ff} = 0.01$ as a fiducial value.

Since $f_{\rm H_2}$ is a function of $\Sigma_g$ alone in our model, we
now have the star formation rate in terms of three parameters:
$\Sigma_g$, $M_{\rm gal}$, and $\rho_{\rm H_2}$. We can eliminate the
last of these on empirical grounds. The Milky Way has $\Sigma_g\approx
100$ $M_\odot$ pc$^{-2}$ and $\rho_{\rm H_2} \approx 100\mu_{\rm H}$
cm$^{-3}$, where $\mu_{\rm H} = 2.34\times 10^{-24}$ g is the mean
mass per H nucleus for gas that is 90\% H and 10\% He by mass
\citep{dobbs14a, heyer15a}. The remaining question is how $\rho_{\rm
  H_2}$ scales as we vary $\Sigma_g$; we assume that it does not vary
with $M_{\rm gal}$ at fixed $\Sigma_g$, since variations of this form
correspond simply to a galaxy having a smaller or larger star-forming
disk. To derive this relationship, we note that observations of
galaxies over a large range in surface densities show that the SFR
surface density is well-correlated with the gas surface density
\citep{kennicutt12a}, $\Sigma_{\rm SFR} \propto \Sigma_g^N$, when the
star formation rate and gas surface densities measured over $\sim 1$
kpc scales. The exact value of the index $N$ is debated in the
literature, and is dependent on the exact sample, fitting method, and
the value assumed to convert CO line luminosity (the most common
method used to measure $\Sigma_g$ in extragalactic observations) to
\htwo \ gas mass
\citep{bigiel08a,blanc09a,narayanan11a,narayanan11b,narayanan12a,
  shetty13b,shetty13a}. We adopt a fiducial value $N=2$, motivated by
theoretical studies that suggest such a relation for LIRGs and ULIRGs
when considering a CO-\htwo \ conversion factor that varies with ISM
physical conditions \citep{narayanan12a}.  Since we also have $\Sigma_{\rm
  SFR} \propto \Sigma_g / t_{\rm ff}$, we immediately have $t_{\rm ff}
\propto \Sigma_g^{1-N}$, and thus $\rho_{\rm H_2} \propto
\Sigma_g^{2(N-1)}$. Combining this scaling with the Milky Way
normalisation described above, we arrive at our fiducial scaling
between $\rho_{\rm H_2}$ and $\Sigma_g$:
\begin{equation}
\label{eq:rhoscaling}
\rho_{\rm H_2} = \rho_{\mathrm{MW}} 
\left(\frac{\Sigma_g}{100\,M_\odot\,\mathrm{pc}^{-2}}\right)^{2(N-1)}
\end{equation}
with $\rho_{\rm MW} = 100 \mu_{\rm H}$ and $N=2$.
We discuss how changing either of the coefficient or index of this relation
would affect our results in \autoref{section:uncertainties}. However, we
note that this scaling produces reasonable values for the Milky Way: the
ISM mass inside the Solar Circle is $M_{\rm gal} \approx
2\times 10^9$ $M_\odot$ considering both
H~\textsc{i} \citep{wolfire03a} and H$_2$ \citep{heyer15a}, and using
$\Sigma_g = 100$ $M_\odot$ pc$^{-2}$ in \autoref{eq:sfr}
gives a total star formation rate of $3.7$ $M_\odot$ yr$^{-1}$, within
a factor of a few of the consensus range of $1-2$ $M_\odot$ yr$^{-1}$
derived by \citet{robitaille10a} and \citet{chomiuk11a}.

We have therefore succeeded in completely specifying our model for the
physical structure of star-forming galaxies and the clouds within them
in terms of two free parameters, $\Sigma_g$ and $M_{\rm gal}$.  We
take to the former be in the range $\log \Sigma_g \in [1.75,3.75]$
$M_\odot$ pc$^{-2}$, and the latter to be in the range $M_{\rm gal} =
10^9 - 10^{10}$ $M_\odot$ for local galaxies, and $\sim
10^{10}-10^{11}$ $M_\odot$ for high redshift ones.  The minimum in the
surface densities is motivated by observations of nearby galaxies
\citep[e.g.][]{bolatto08a,leroy13a}, while the range in galaxy gas
masses is constrained by surveys of galaxies near and far
\citep{saintonge11a,bothwell13a}.

Finally, we convert between star formation rate and observed infrared luminosity
employing the \citet{murphy11a} conversion as
summarised by \citet{kennicutt12a},
\begin{equation}
\log_{10}(L_{\rm IR} (3-1100\, \mu{\rm m})) = \log_{10}(\mathrm{SFR})+43.41.
\end{equation}

\subsection{Chemical Structure}
\label{section:chemistry}

As mentioned above, our clouds consist of radial layers, each
chemically independent from one another. Each cloud contains
$N_{\rm zone}$ zones, with a default $N_{\rm zone} = 16$. We
show in \S~\ref{section:resolution} that this is sufficient to produce a
converged result. We assign each cloud a center-to-edge column density
of H nuclei $N_{\rm H} = (3/4) \Sigma_g/\mu_{\rm H}$; the factor
of $(3/4)$ is the simply the difference between the mean column
density and the center-to-edge column density. In this model,
we assign the $i$th zone (starting with $i=0$) to cover the range
of column densities from $[i/N_{\rm zone}] N_{\rm H}$ to
$[(i+1)/N_{\rm zone}] N_{\rm H}$, with a mean column density
$N_{{\rm H},i} = [(i+1/2)/N_{\rm zone}] N_{\rm H}$. We
calculate the mass in each zone from the column density
range and the volume density, assuming spherical geometry.

In each zone we must determine the chemical state of the carbon and
oxygen atoms. These two species, either separately in atomic form or
combined into CO, are the dominant coolants and line emitters.  We
adopt total abundances of $[{\rm C}/{\rm H}] = 2\times 10^{-4}$ and
$[{\rm O}/{\rm H}] = 4\times 10^{-4}$ for C and O, respectively,
consistent with their abundances in the Milky Way \citep{draine11a}.
In principle, depending on the physical properties of the gas in each
layer, the carbon can be stored predominantly as C$^+$, C, or CO.
Similarly, the hydrogen can be in atomic or molecular form in any
given layer, and this chemical state of the H in turn affects that of
the C and O. To model these effects, we compute the chemical state of
each zone using the reduced carbon-oxygen chemical network developed
by \citet{nelson99a} combined with the \citet{glover07a}
non-equilibrium hydrogen chemical network,\footnote{By using the
  \citet{glover07a} model for hydrogen chemistry, we are folding an
  inconsistency into our model.  Specifically, to determine the bulk
  physical properties of our model clouds, including the star
  formation rates, we utilise the KMT model to decompose the ISM
  phases into neutral hydrogen and molecular.  This is unavoidable as
  the networks require knowledge of the background radiation field
  and cosmic ray ionisation rate, both of which likely depend on the
  star formation rate and thus the \htwo \ fraction. To be fully
  consistent we would be required to iterate between chemistry and
  star formation, which would be quite computationally
  expensive. Hereafter in the paper, all hydrogen phase abundances
  that we quote derive from what is explicitly calculated in the
  chemical equilibrium network. However, the level of inconsistency is
  generally small, in that the H$_2$ fraction that results from the
  explicit chemical modeling never deviates that strongly from the KMT
  prediction, varying by a maximum of a factor of $\sim 2$ within the
  surface density range of interest.}  combined following the procedure
  described in \citet{glover12a}.  We summarise the reactions included
  in our network, and the rate coefficients we use for them, in
  \autoref{table:chemistry}. We refer readers to \citet{glover12a} for
  full details on the network and its implementation, but we mention
  here three choices that are specific to the model we use in this
  paper.

\begin{table*}
\centering
\caption{
\label{table:chemistry}
Coefficients adopted for our chemical network, following
\citet{glover12a}.  In this table, $\zeta$ is the cosmic ray primary
ionisation rate, $\chi_{\rm FUV}$ is the normalized FUV radiation
field strength, $A_V$ is the visual extinction, $x_{\rm H} = n_{\rm
  H}/n$ is the H abundance, $x_{\rm H_2} = n_{\rm H_2}/n$ is the H$_2$
abundance, $T_4 = T/(10^4\,\mathrm{K})$, \lnte = ln$(8.6173 \times
10^{-5} \times T/\mathrm{K})$.  Note that this network includes
several super-species: CH$_x$ agglomerates CH, CH$_2$, etc., and
similarly for OH$_x$, and M and M$^+$ agglomerate a number of metallic
species with low ionisation potentials (e.g., Fe, Si).}
\begin{tabular}{ll}
\hline Reaction & Rate Coefficient \\
\hline
Cosmic Ray Reactions [s$^{-1}$ molecule$^{-1}$]: & \\
cr + H $\rightarrow$ H$^+$ + e  & $\zeta$  \\
cr + H$_2 \rightarrow$ H$_3^+$ + e + H + cr & $2\zeta$  \\
cr + He $\rightarrow$ He$^+$ + e + cr & $1.1\zeta$ \\
\hline
Photoreactions [s$^{-1}$ molecule$^{-1}$]: & \\
$\gamma$ + H$_2$ $\rightarrow$ 2 H & $5.6 \times 10^{-11} \chi_{\mathrm{FUV}} f_{\mathrm{shield}}(N_{\rm \htwo}) e^{-3.74 A_V}$ \\
$\gamma$ + CO $\rightarrow$ C + O & $2 \times 10^{-10} \chi_{\mathrm{FUV}} f_{\mathrm{shield}}(N_{\rm CO},N_{\rm \htwo}) e^{-3.53 A_V}$ \\
$\gamma$ + C $\rightarrow$ C$^+$ + e & $3 \times 10^{-10} \chi_{\mathrm{FUV}} e^{-3 A_V}$ \\
$\gamma$ + CH$_x$ $\rightarrow$ C + H & $1 \times 10^{-9} \chi_{\mathrm{FUV}} e^{-1.5 A_V}$ \\
$\gamma$ + OH$_x$ $\rightarrow$ O + H & $5 \times 10^{-10} \chi_{\mathrm{FUV}} e^{-1.7 A_V}$ \\
$\gamma$ + M $\rightarrow$ M$^+$ + e & $3.4 \times 10^{-10} \chi_{\mathrm{FUV}} e^{-1.9 A_V}$ \\
$\gamma$ + HCO$^+$ $\rightarrow$ CO + H & $1.5 \times 10^{-10} \chi_{\mathrm{FUV}} e^{-2.5 A_V}$ \\
\hline
Ion-Neutral Reactions [cm$^3$ s$^{-1}$ molecule$^{-1}$]: &\\
H$_3^+$ + CI $\rightarrow$ \chx + \htwo & $2 \times 10^{-9}$ \\
H$_3^+$ + OI $\rightarrow$ \ohx + \htwo & $8 \times 10^{-10}$ \\
H$_3^+$ + CO $\rightarrow$ HCO$^+$ + \htwo & $1.7 \times 10^{-9}$ \\
He$^+$ + \htwo $\rightarrow$ He + H + H$^+$ & $7 \times 10^{-15}$ \\
He$^+$ + CO $\rightarrow$ C$^+$ + O + He& $1.4 \times 10^{-9}/\sqrt{T/300}$ \\
\cp + \htwo $\rightarrow$ \chx + H & $4 \times 10^{-16}$ \\
\cp + \ohx $\rightarrow$ HCO$^+$ & $1 \times 10^{-9}$ \\
\hline
Neutral-Neutral Reactions [cm$^3$ s$^{-1}$ molecule$^{-1}$]: &\\
OI + \chx $\rightarrow$ CO + H & $2 \times 10^{-10}$ \\
CI + \ohx $\rightarrow$ CO + H & $5 \times 10^{-12} \sqrt{T}$ \\
\hline
Recombinations and Charge-Transfers [cm$^3$ s$^{-1}$ molecule$^{-1}$]:&\\
\hep + e $\rightarrow$ He + $\gamma$ & $1 \times 10^{-11}/\sqrt{T} \times $\\
&$(11.19-1.676 \times {\rm log_{10}}(T) - 0.2852 \times {\rm log_{10}}(T^2) + 0.04433\times{\rm log_{10}}(T^3))$ \\
H$_3^+$ + e $\rightarrow$ \htwo + H & $2.34 \times 10^{-8} (T/300)^{-0.52}$ \\
H$_3^+$ + e $\rightarrow$ 3H & $4.36\times 10^{-8} (T/300)^{-0.52}$ \\
\cp + e $\rightarrow$ CI + $\gamma$ & $4.67 \times 10^{-12} (T/300)^{-0.6}$ \\
HCO$^+$ + e $\rightarrow$ CO + H & $2.76 \times 10^{-7} (T/300)^{-0.64}$ \\
M$^+$ + e $\rightarrow$ M + $\gamma$ & $3.8 \times 10^{-10} T^{-0.65}$ \\
H$_3^+$ + M $\rightarrow$ M + $\gamma$ & $2 \times 10^{-9}$ \\
\hline
Hydrogenic Chemistry [cm$^3$ s$^{-1}$ molecule$^{-1}$]:&\\
H$^+$ + e $\rightarrow$ H & $2.753 \times 10^{-14} \times (315614/T)^{1.5} \times (1+(115188/T)^{0.407})^{-2.242}$\\
\htwo + H $\rightarrow$ 3H & \khl$ = 6.67 \times 10^{-12} \sqrt{T} \times e^{-\left(1.0 + \frac{63590}{T}\right)}$ \\
& \khh$ = 3.52\times 10^{-9} e^{\frac{-43900}{T}}$ \\
& $n_{\rm cr,H} = 10^{\left(3 - 0.416 \times {\rm log_{10}}T_4 - 0.327\times ({\rm log_{10}}(T_4))^2  \right)}$ \\
& $n_{\rm cr,H2} = 10^{\left( 4.845 - 1.3 \times {\rm log_{10}}(T_4) + 1.62 \times ({\rm log_{10}}(T_4))^2  \right)}$\\ 
& $n_{\rm cr} = \left( x_{\rm H}/n_{\rm cr,H} + x_{\rm H2}/n_{\rm cr,H2}\right) ^{-1}$\\
& ${\rm exp}\left[ (n/n_{\rm cr})/(1.+n/n_{\rm cr}) \times {\rm log_{10}}(\khh) + 1/(1+n/n_{\rm cr}) \times {\rm log_{10}}(\khl)\right]$\\
\htwo + \htwo $\rightarrow$ \htwo + 2H &  $\khtwol  = 5.996\times 10^{-30} T^{4.1881}/\left( 1+6.761\times 10^{-6} \times T \right)^{(5.6881)} \times e^{-54657.4/T}$\\ 
& \khtwoh $ = 1.9 \times 10^{-9} \times e^{-53300/T}$ \\
& ${\rm exp}\left[(n/n_{\rm cr})/(1.+(n/n_{\rm cr})) \times {\rm log_{10}}(\khtwoh) + 1/(1+n/n_{\rm cr}) \times {\rm log_{10}}(\khtwol)\right]$\\
H + e $\rightarrow$ H$^+$ + 2e& ${\rm exp}[ -37.7 + 13.5 \times \lnte - 5.7 \times \lnte^2 + 1.6 \times \lnte^3 - $\\
&$ 0.28 \times \lnte^4 + 0.03\times \lnte^5 - 2.6\times \lnte^6 + 1.1 \times \lnte^7 - 2.0 \times \lnte^8]$\\
He$^+$ + \htwo $\rightarrow$ H$^+$ + He + H& $3.74 \times 10^{-14} e^{-35/T}$\\
H + H + grain $\rightarrow$ \htwo + grain & $f_A = 1/(1+10^4 \times e^{-600/T_d})$\\
& $3 \times 10^{-18} \sqrt{T} * f_A / (1+0.04 * \sqrt{T+T_d} + 0.002 T + 8\times10^{-6} \times \sqrt{T})$\\
H$^+$ + e + grain $\rightarrow$ H + grain& $\psi = \chi \sqrt{T}/n_e$\\
& $12.25 \times 10^{-14} / \left(1 + 8 \times 10^{-6} \psi^{1.378}  \left(1 + 508 \times T^{0.016}\psi^{-0.47-1.1\times 10^{-5} {\rm ln}(T)}\right) \right)$\\
\hline
\end{tabular}
\end{table*}

First, the network requires that we specify the strength of the
unshielded interstellar radiation field (ISRF). We
characterise this in terms of the FUV radiation intensity
normalised to the Solar neighborhood value, $\chi_{\rm FUV}$. We
assume that $\chi_{\rm FUV}$ is proportional to a galaxy's SFR
normalised to the $1$ $M_\odot$ yr$^{-1}$ SFR of the Milky Way
\citep{robitaille10a, chomiuk11a}: $\chi_{\rm FUV} = \mbox{SFR} /
(M_\odot\,{\rm yr}^{-1})$.

Second, we must compute the amount by which all photochemical
reaction rates are reduced in the interiors of clouds by shielding
of the ISRF. The \despotic~implementation of the \citet{glover12a}
network that we use includes reductions in the rates of all
photochemical reactions by dust shielding, and
reductions in the rates of H$_2$ and CO dissociation by self-shielding
and (for CO) H$_2$ cross shielding. We characterise dust shielding in
terms of the visual extinction $A_V = (1/2) (A_V/N_{\rm H}) N_{\rm H}$,
where the ratio $(A_V/N_{\rm H})$ is the dust extinction per H
nucleus at V band (\autoref{table:free_parameters}). The factor of $(1/2)$ gives
a rough average column density over the volume of the cloud. We
evaluate the reduction in the H$_2$ dissociation rate using the
shielding function of \citet{draine96a}, which is a function of
the H$_2$ column density and velocity dispersion; for the latter
we use the value given by \autoref{eq:sigmac}, while for the
former we use $N_{\rm H_2} = x_{\rm H_2} N_{\rm H}$, where $x_{\rm H_2}$
is the abundance of H$_2$ molecules per H nucleus in the zone
in question; note that each zone is independent, so we do not use
information on the chemical composition of outer zones to evaluate
$x_{\rm H_2}$, a minor inconsistency in our model. Similarly, we
compute the reduction in the CO photodissociation rate using an
interpolated version of the shielding function tabulated by
\citet{vandishoeck88a}, which depends on the CO and H$_2$ 
column densities. We evaluate the CO column density as $N_{\rm CO}
= x_{\rm CO} N_{\rm H}$, in analogy with our treatment of the H$_2$
column. We determine both abundances $x_{\rm H_2}$ and $x_{\rm CO}$
by iterating the network to convergence -- see \autoref{ssec:convergence}.

Third, we must also specify the cosmic ray primary ionisation rate
$\zeta$.  The value of this parameter even in the Milky Way is
significantly uncertain.  Recent observations suggest a value $\zeta
\sim 10^{-16}$ s$^{-1}$ in the diffuse ISM \citep{neufeld10a,
  indriolo12a}, but if a significant amount of the cosmic ray flux is
at low energies, the ionisation rate in the interiors of molecular
clouds will be lower due to shielding; indeed, a rate as high as
$10^{-16}$ s$^{-1}$ appears difficult to reconcile with the observed
low temperatures of $\sim 10$ K typically found in molecular gas
\citep{narayanan12a,narayanan12b}. For this reason, we adopt a more
conservative value of $\zeta = 10^{-17}$ s$^{-1}$ as our fiducial
choice for the Milky Way. We discuss how this choice influences our
results in \autoref{section:theory}. We further assume that the cosmic
ray ionisation rate scales linearly with the total star formation rate
of a galaxy, so our final scaling is $\zeta_{-16} = 0.1\times
\mbox{SFR}/(M_\odot\,\mathrm{yr}^{-1})$, where $\zeta_{-16} \equiv
\zeta/10^{-16}$ s$^{-1}$. Note that this choice of scaling too is
significantly uncertain, and others are plausible.\footnote{For
  example, \citet{papadopoulos10a} and \citet{bisbas15a} assume that
  cosmic ray intensity scales as the volume density of star formation
  rather than the total rate of star formation; which assumption is
  closer to reality depends on the extent to which cosmic rays are
  confined by magnetic fields and subject to losses as they propagate
  through a galaxy.}

\subsection{Thermal State of Clouds}
\label{ssec:thermal_struct}

The third component of our model is a calculation of the gas
temperature, which we compute independently for each zone of our
model clouds. We find the temperature by balancing the relevant
heating and cooling processes, as well as energy exchange with dust.
Following \citet{goldsmith01a}, the processes we consider are
photoelectric and cosmic ray heating of the gas, line cooling of the
gas by C$^+$, C, O, and CO, heating of the dust by the ISRF and by a
thermal infrared field, cooling of the dust by thermal emission, and
collisional exchange between the dust and gas. We also include cooling
by atomic hydrogen excited by electrons via the Lyman $\alpha$ and
Lyman $\beta$ lines and the two-photon continuum, using interpolated
collisional excitation rate coefficients \citep[Table 3.16]{osterbrock06a};
these processes become important at temperatures above $\sim 5000$ K,
which are sometimes reached in the outer zones of our clouds.
Formally,
\begin{eqnarray}
\label{equation:goldsmith1}
\Gamma_{\rm pe} + \Gamma_{\rm CR} - \Lambda_{\rm line} -\Lambda_{\rm H} + \Psi_{\rm gd} &  =  &0\\
\label{equation:goldsmith2}
\Gamma_{\rm ISRF} + \Gamma_{\rm thermal} - \Lambda_{\rm thermal} - \Psi_{\rm gd} & = & 0. 
\end{eqnarray}
Terms denoted by $\Gamma$ are heating terms, those denoted by $\Lambda$
are cooling terms, while the gas-dust energy exchange term $\Psi_{\rm gd}$ can
have either sign depending on the gas-dust temperature difference; our convention
is that a positive sign corresponds to dust being hotter than the gas, leading to a
transfer from dust to gas.

As with the chemical calculation, we solve these equations
using the \despotic~code, and we refer readers to \citet{krumholz13a} for a full
description of how the rates for each of these processes are computed.
The parameters we adopt are as shown in \autoref{table:free_parameters}.
Note that the line cooling rate depends on the statistical equilibrium calculated
as described in \autoref{section:radiative_transfer}.

\subsection{Statistical Equilibrium}
\label{section:radiative_transfer}

The final part of our model is statistical equilibrium within the
level populations of each species. The \despotic~code computes these
using the escape probability approximation for the radiative transfer
problem. Formally we determine the fraction $f_i$ of each species in
quantum state $i$ by solving the linear system
\begin{eqnarray}
\lefteqn{\sum_j f_j \left[q_{ji} + \beta_{ji} (1 + n_{\gamma,ji}) A_{ji} + 
\beta_{ij} \frac{g_i}{g_j} n_{\gamma,ij} A_{ij} \right]}
\nonumber
\\
& = & f_i \sum_k \left[q_{ik} + \beta_{ik} (1 + n_{\gamma,ik}) A_{ik} + 
\beta_{ki} \frac{g_k}{g_i} n_{\gamma,k i} A_{ki}\right]\qquad
\label{equation:kmt_stateq}
\end{eqnarray}
subject to the constraint $\sum_i f_i = 1$.
Here $A_{ij}$ is the Einstein coefficient for spontaneous transitions from state
$i$ to state $j$, $g_i$ and $g_j$ are the degeneracies of the states,
\begin{equation}
n_{\gamma,ij} = \frac{1}{\exp(\Delta E_{ij}/k_B T_{\rm CMB})-1}
\end{equation}
is the photon occupation number of the cosmic microwave background at
the frequency corresponding to the transition between the states,
$E_{ij}$ is the energy difference between the states, and $\beta_{ij}$
is the escape probability for photons of this energy. We compute the
escape probability for each shell independently, assuming a spherical
geometry.  The escape probabilities computed include the effects of
both resonant and dust absorption -- see \citet{krumholz13a} for
details. Finally, $q_{ij}$ is the collisional transition rate between
the states, which is given by $q_{ij} = f_{\rm cl} n_{\rm H}
k_{\mathrm{H},ij}$ or $q_{ij} = f_{\rm cl} n_{\rm H_2}
k_{\mathrm{H_2},ij}$ in the H~\textsc{i} and H$_2$ regions,
respectively; the quantities $k_{\mathrm{H},ij}$ and
$k_{\mathrm{H_2},ij}$ are the collision rate coefficients, $n_{\rm H}$
and $n_{\rm H_2}$ are the number densities of H atoms or H$_2$
molecules, and $f_{\rm cl}$ is a factor that accounts for the
enhancement in collision rates induced by turbulent clumping. 

All the Einstein coefficients and collisional rate coefficients required for our
calculation come from the Leiden Atomic and Molecular Database \citep{schoier05a}.
In particular, we make use of the following collision rate coefficients: C$^+$ with
H \citep{launay77a, barinovs05a}, C$^+$ with H$_2$ \citep{lique13a, wiesenfeld14a},
C with H \citep{launay77a}, C with He \citep{staemmler91a}, C with H$_2$
\citep{schroder91a}, O with H \citep{abrahamsson07a}, O with H$_2$ \citep{jaquet92a},
and CO with H$_2$ \citep{yang10a}.

\subsection{Convergence and Computation of the Emergent Luminosity}
\label{ssec:convergence}

Calculation of the full model proceeds via the following steps. First,
we compute the physical properties of each cloud following the method
given in \autoref{ssec:phys_struct}. Armed with these, we guess an
initial temperature, chemical state, and set of level populations for
each layer in the cloud. We then perform a triple-iteration procedure,
independently for each zone. The outermost
loop is to run the chemical network (\autoref{section:chemistry}) to
convergence while holding the temperature fixed. The middle loop is to
compute the temperature holding the level populations fixed
(\autoref{ssec:thermal_struct}). The innermost loop is to iterate the
level populations of each species to convergence
(\autoref{section:radiative_transfer}). We iterate in this manner
until all three quantities -- chemical abundances, temperature, and
level populations -- remain fixed to within a certain tolerance, at
which point we have found a consistent chemical, thermal, and
statistical state for each zone.

Once the level populations are in hand, it is straightforward to compute the
observable luminosity in the \cii~158 $\mu$m line, or in any other
transition. The total luminosity per unit mass produced in a line produced
by molecules or atoms of species $S$ transitioning between states $i$ 
and $j$, summed over each zone, is
\begin{equation}
L_{ij}/M = \mu_{\rm H}^{-1} x_S \beta_{ij} \left[\left(1 + n_{\gamma,ij}\right) f_i
- \frac{g_i}{g_j} n_{\gamma,ij} f_j\right] A_{ij} E_{ij},
\end{equation}
where $x_S$ is the abundance of the species and $f_i$ and $\beta_{ij}$ are
the level populations and escape probabilities in each layer. Each
zone $n$ has a mass $M_n$, computed from its range of column
densities, and the total luminosity of the $N_c$ clouds in the entire 
galaxy is simply
\begin{eqnarray}
\lefteqn{
L_{ij} = 
N_c  \mu_{\rm H}^{-1} A_{ij} E_{ij} \sum_n M_n x_{S,n} 
\beta_{ij,n} \cdot {}
}
\nonumber \\
& & \qquad \left[\left(1 + n_{\gamma,ij}\right) f_{i,n}
- \frac{g_i}{g_j} n_{\gamma,ij} f_{j,n}\right],
\end{eqnarray}
where $x_{S,n}$, $f_{i,n}$, and $\beta_{ij,n}$ are the abundance,
level population fraction, and escape probability in the $n$th zone
of our model clouds.

\subsection{Sample Results}

\begin{figure*}
\hspace{-0.75cm}

\includegraphics[scale=0.75]{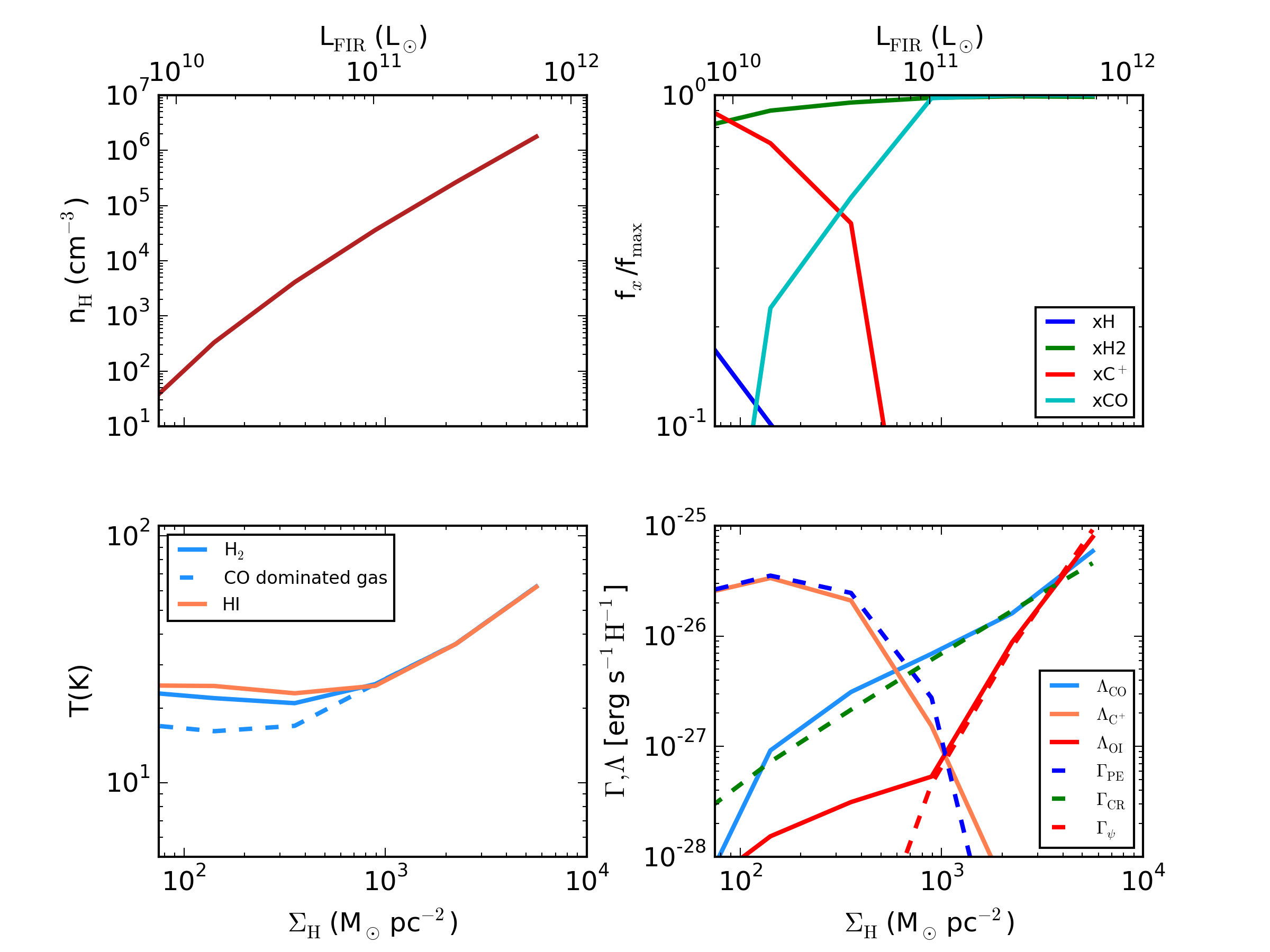}
\caption{Physical properties of model clouds inside a galaxy of
$M_{\rm gas} = 10^9$ \msun as a function of cloud surface
density.  These include volume density (top left), chemical
abundances (top right), gas kinetic temperature (bottom left), and
heating/cooling rates (bottom right). All quantities shown are
mass-weighted averages, with the exception that In the gas kinetic temperature
panel the different lines correspond to temperatures weighted by H~\textsc{i}
mass, H$_2$ mass, and CO luminosity, respectively.
\label{figure:physical_properties_sd}}
\end{figure*}

Before moving on to our results for \cii\ emission, in this section we
provides a brief example of the thermal and chemical properties that
our models produce. These will provide the reader with some intuition
for how the physical, thermal, and chemical properties of our model
clouds vary with galaxy infrared luminosity, or mean cloud surface
density. For this example, we consider a galaxy with a gas mass of
$M_{\rm gas} = 10^9$ $M_\odot$ (i.e. similar to the Milky Way), and we
vary the surface density $\Sigma_g$ within the range specified in
\autoref{table:free_parameters}. For each value of $\Sigma_g$ we
derive a star formation rate (\autoref{eq:sfr}) and thus a far
infrared luminosity, and we run the chemical-thermal-statistical
network to equilibrium following the procedure described in
\autoref{ssec:convergence}. We summarise the resulting cloud
properties as a function of FIR luminosity and $\Sigma_{\rm g}$ in
\autoref{figure:physical_properties_sd}, where we show the cloud mean
densities, fractional chemical abundances for a few species, gas
kinetic temperatures, and ISM heating/cooling rates as a function of
cloud surface density (and galaxy infrared luminosity).

The fractional abundance subpanels of \autoref{figure:physical_properties_sd}
summarise the central arguments
laid out in this paper.  As the total cloud surface densities rise, so
does the typical mass fraction of gas in the \htwo \ phase
\citep{krumholz08a}, owing to the increased ability of hydrogen to
self-shield against dissociating Lyman-Werner band photons.  This
point is especially pertinent to our central argument.  At a fixed
galaxy mass, increased gas surface densities lead to increased star
formation rates.  In these conditions, the molecular to atomic ratio
in giant clouds increases.  At the same time, with increasing cloud
surface density (or galaxy star formation rate), \cp \ abundances
decline, and CO abundances increase.  This owes principally to the
role of dust column shielding CO from photodissociating radiation.  The
fraction of the cloud that is dominated by \cp \ hence decreases with
increasing cloud surface density.

  For the temperature, we discriminate between \htwo \ and \hi \ gas
  in clouds, and plot the mass-weighted values for each phase.  We
  additionally show the CO luminosity-weighted gas temperature with
  the dashed line, as this is the temperature that most closely
  corresponds to observations.  At low cloud surface densities, the CO
  dominates the cool ($T \sim 15 $ K) inner parts of clouds, though
  the warmer outer layers are dominated by \ci \ and \cp. The bulk of
  the mass is at these warmer temperatures, and the heating is
  dominated by the grain photoelectric effect.  At higher cloud
  surface densities, the bulk of the cloud is dominated by CO.  Here, the
  grain photoelectric effect is less effective owing to increased
  $A_{\rm V}$, but the impact of cosmic ray ionisations and energy
  exchange with dust is increased.

\section{Results}
\label{section:results}
In the model that we develop, the \cii-FIR deficit in galaxies
principally owes to a combination of atomic PDRs serving as the
dominant site of \cp \ in galaxies, and a decreasing atomic to
molecular fraction in galaxies of increasing luminosity.  In this
section, we lay the case for this argument in detail.

\subsection{Carbon-based Chemistry}
\label{section:chemistry_intuition}
\begin{figure}
\hspace{-0.75cm}
\includegraphics[scale=0.45]{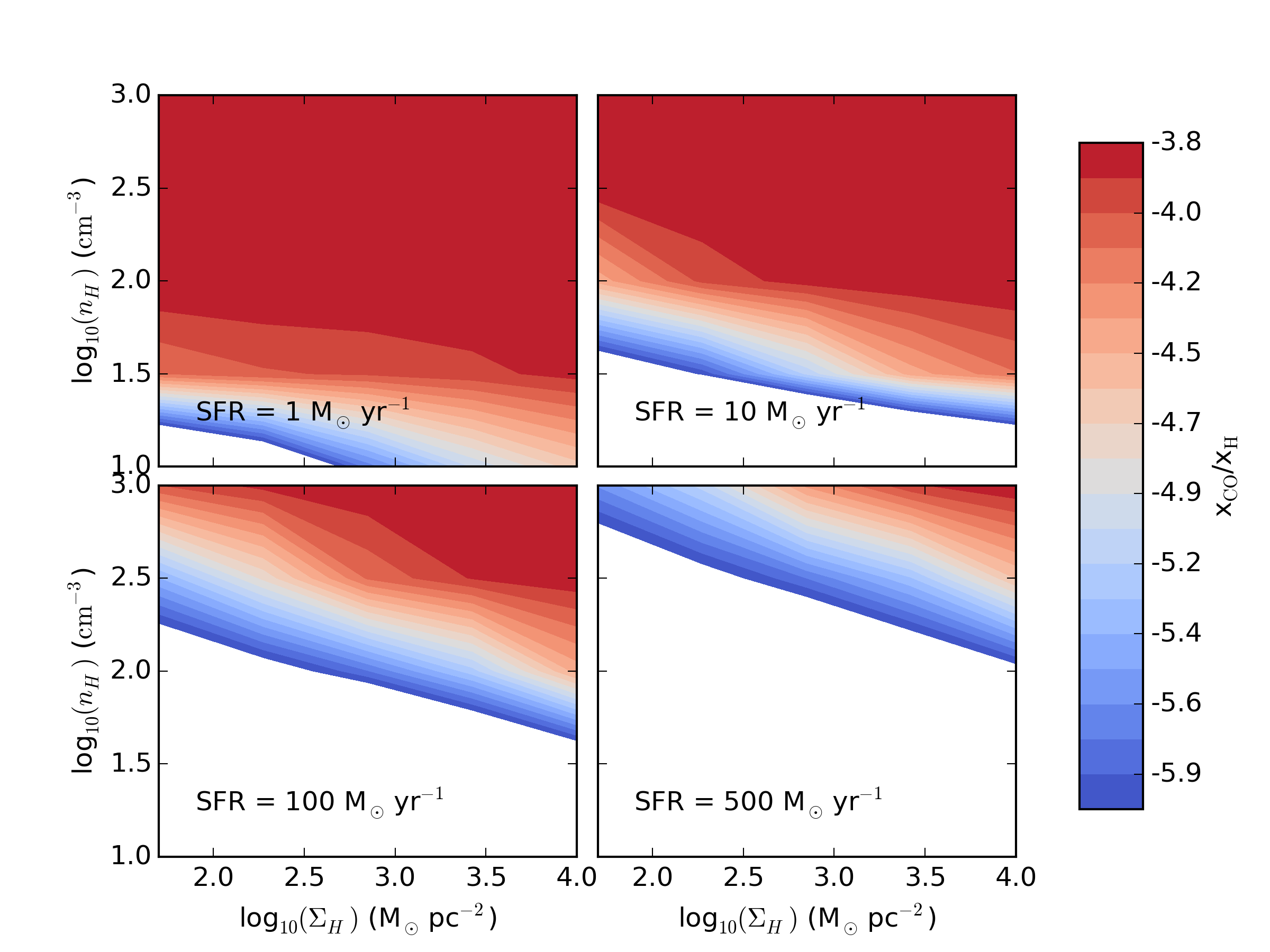}
\caption{$^{12}$CO abundance contours as a function of the volumetric
  and gas surface densities.  Abundance contours are shown for four
  fiducial star formation rates, which control the far ultraviolet
  flux and the cosmic ray ionisation rates.  At low radiation field
  strengths/cosmic ray ionisation rates, nearly all of the carbon is
  in the form of CO.  At higher SFRs, UV radiation and cosmic ray
  ionisations contribute to the destruction of CO, especially at low
  densities and surface densities.  The whitespace in the contour
  subplots denotes abundances below $10^{-6}$/\htwo.  \label{figure:co_contours}}
\end{figure}

We begin with the instructive question: how is CO in clouds typically
formed and destroyed?  The principle formation channels for CO are via
neutral-neutral reactions with \chx \ and \ohx.
CO is destroyed in the ISM via both cosmic rays and ultraviolet
radiation.  The molecule is directly destroyed most efficiently via
interactions with ionised helium, \hep, which is created via cosmic
ray ionisations of neutral He.  At the same time, FUV radiation can
also reduce CO abundances via a variety of channels: it can directly
destroy CO, as well as prevent its formation via the photodissociation
of CO's main formation reactants, \ci, \chx, and \ohx.  Once carbon is
in neutral atomic form, UV radiation can ionise \ci \ in order to form
\cp.

Opposing CO dissociation and ionisation (and consequently the formation of
\cp) are the surface density and volumetric density of the cloud.  To
understand the role of the surface density, consider the
photo-reactions in \autoref{table:chemistry}.  The dissociation rates
are only linearly dependent on the ultraviolet radiation field
strength, but exponentially decrease with increased $A_V$.  In
particular, increased surface densities prevent the dissociation and
ionisation of \chx, \ohx, and CO molecules, as well as \ci.

Increased volumetric densities, $n_{\rm H}$, also promote neutral atom and
molecule formation, and prevent the formation of \cp.  Again, consider
the photoionisation of neutral carbon.  The reaction rates for
photodissociations and photoionisations are density-independent.
However, the recombination and molecular formation rates within the
ion-molecule, ion-atom and neutral-neutral reactions all scale
linearly with density.  Hence, given sufficient density, recombination
and molecule formation outpace the ionisation rates.

The carbon-based chemistry in clouds in galaxies is therefore set by a
competition between the star formation rate of the galaxy, and the
density and surface density of clouds.  The star formation rate
controls the cosmic ray ionisation rate, as well as the ultraviolet
flux.  As a result, all else being equal, increased SFRs result in
decreased molecular CO abundances, and increased \ci \  and \cp
\ abundances.

\begin{figure}
\hspace{-0.75cm}
\includegraphics[scale=0.45]{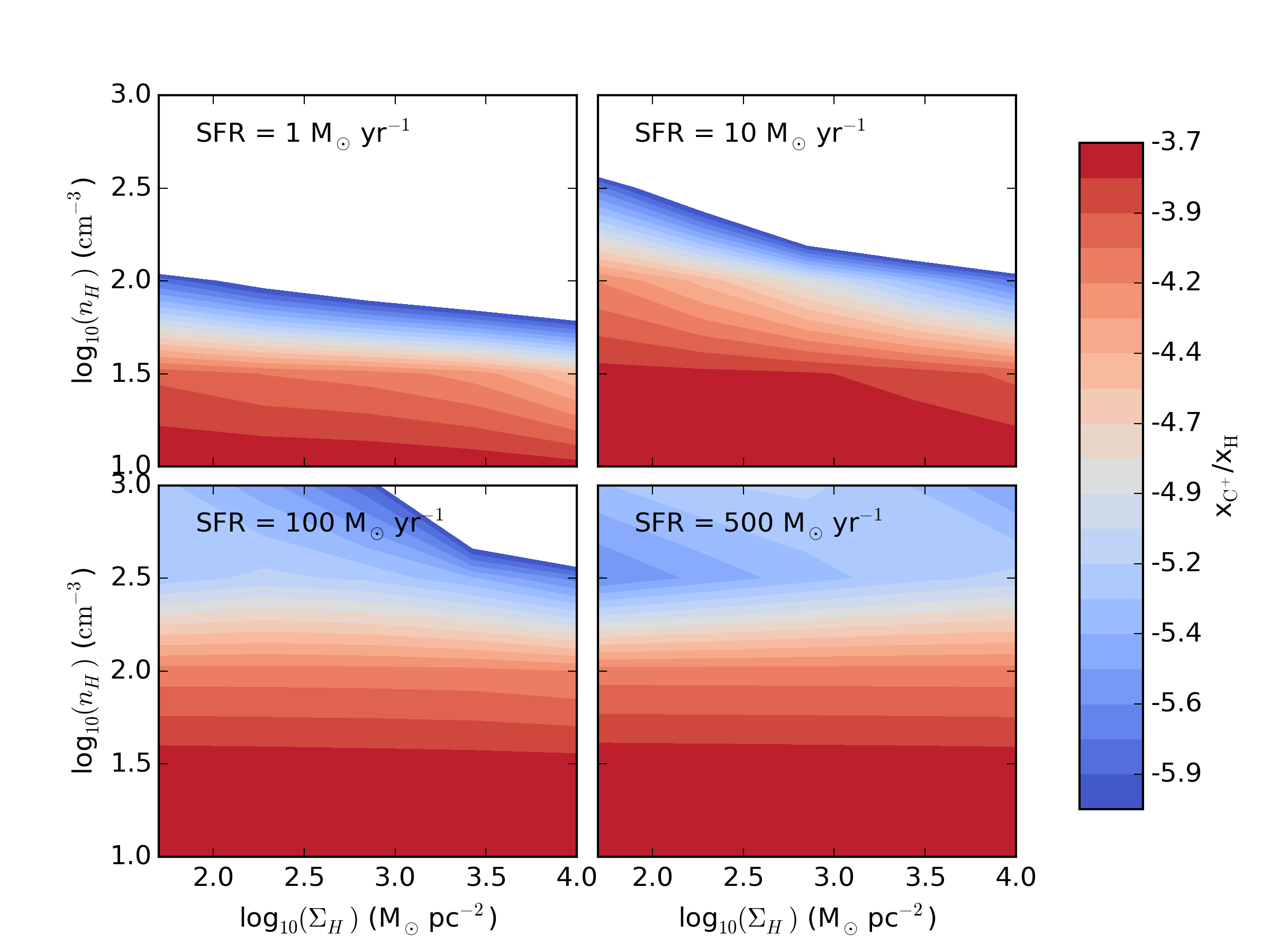}
\caption{Similar to Figure~\ref{figure:co_contours}, but \cp
  \ abundance contours as a function of gas surface density and
  volumetric density.  \cp \ abundances are increased in low volume
  density and surface density gas, and in high ultraviolet radiation
  fields and/or cosmic ray fluxes. The whitespace in the contour
  subplots denotes abundances below
  $10^{-6}$/\htwo. \label{figure:cii_contours}}
\end{figure}

In order to provide the reader with some intuition as to how these
effects drive carbon-based chemistry, in
Figures~\ref{figure:co_contours} - \ref{figure:cii_contours}, we show
the CO and \cp \ abundances for a grid of model cloud densities and
surface densities given a range of star formation rates (and FUV
fluxes and cosmic ray ionisation rates that scale, accordingly).  Note
that these cloud models are principally for the purposes of the
illustration of dominant physical effects, and therefore have not been
constructed via the methods in \autoref{section:model} (meaning that
the SFR, $\Sigma_{\rm SFR}$, densities and surface densities are not
all interconnected, nor is there a multiphase breakdown of these
clouds; they are of a single ISM phase).  The effects of increased
cosmic ray fluxes and FUV radiation field strengths via increased SFR
on the carbon-based chemistry is clear.  At low star formation rates,
even relatively low $n_{\rm H}$ and $\Sigma_{\rm H}$ gas is sufficiently
well-shielded that the carbon can exist in molecular (CO) form.
Consequently, \cp \ is confined to the most diffuse gas at low SFRs.
At higher SFRs the situation is reversed.  Increased $\chi_{\rm FUV}$ and cosmic
ray ionisation rates dissociate CO and ionise C, thereby increasing
\cp \ abundances in clouds with a large range in physical conditions.
\cp \ is destroyed, and \ci \ and CO are most efficiently formed, when the
volume density and surface density of the cloud simultaneously
increase.

The numerical experiments represented in
Figures~\ref{figure:co_contours} - \ref{figure:cii_contours} give some
intuition as to how carbon will behave in different physical
environments.  In the remainder of this paper, we build upon this by
combining this with our model for clouds in galaxies developed in
\autoref{section:model}.

\subsection{Application to Multiphase Clouds}
\begin{figure*}
  \centering
\includegraphics[scale=1]{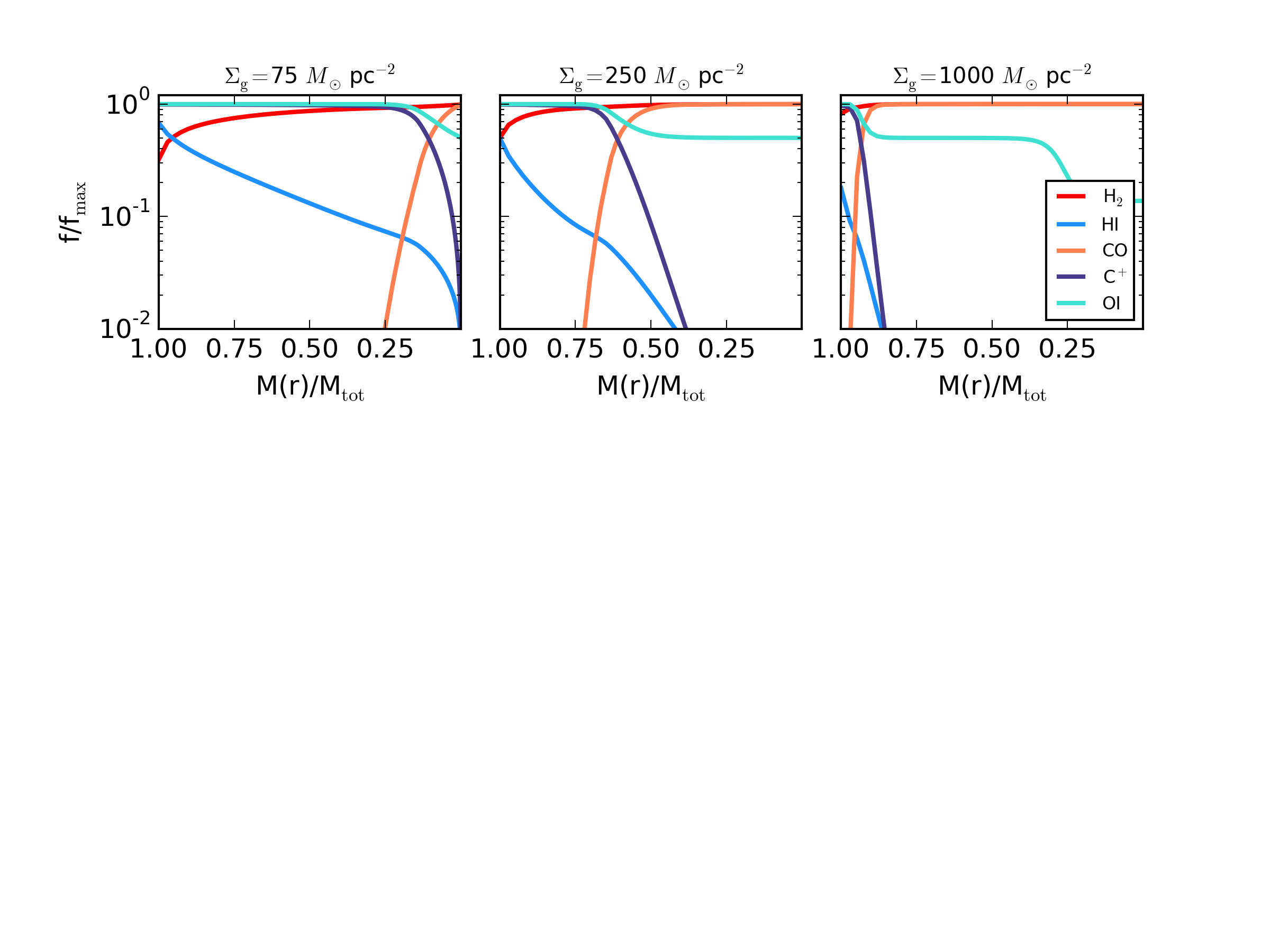}
\vspace{-8.5cm}
\caption{Radial variation of atomic and molecular abundances in three
  model clouds in a galaxy of $M_{\rm gas} = 10^9$ \msunend.  The
  abundances are plotted against the Lagrangian mass enclosed such
  that a value of M(r)/M$_{\rm tot}$ = 1 corresponds to the surface of
  the cloud.  Abundances are normalised to their maximum possible
  value, and plotted as a function of cloud radius.  We show three
  clouds of increasing gas surface density.  In general terms, the
  \htwo \ abundances increase toward the interior of clouds as
  shielding protects molecules from photodissociation.  The carbon
  transitions from \cp \ to CO, with the \cp \ predominantly residing
  in the atomic PDRs and outer shell of \htwo \ gas.  Clouds of
  increasing surface density have an increasing fraction of their gas
  in molecular \htwo \ form, and increasing fraction of their carbon
  locked in CO molecules.  The oxygen abundances decrease toward the
  interior of clouds as the atom becomes locked up in CO and,
  in the inner parts of the highest surface density case, \ohx
  \ molecules.
\label{figure:abundances}}
\end{figure*}

We are now in a place to understand the fractional abundances of
carbon in its different phases in giant clouds.  In
\autoref{figure:abundances}, we present the radial fractional
abundances for a variety of relevant species in our chemical reaction
networks for three clouds of increasing surface density for a galaxy
with gas mass $10^9$ \msunend.  These clouds are
created within the context of the physical models developed in
\autoref{ssec:phys_struct}, and therefore have increased star formation
rates (and UV fluxes/cosmic ray ionisations) with increasing cloud
surface density.

For low surface density clouds, the hydrogen toward the outer most
layers of the clouds is in atomic form.  In these low surface density
layers, photodissociation destroys \htwo, forming a photodissociated
region (PDR) layer.  At increasing cloud depths and surface densities,
shielding by both gas and dust protects the gas from photodissocation,
and hydrogen can transition from atomic to molecular phase via
grain-assisted reactions.

The carbon chemistry follows a similar broad trend as the hydrogen
chemistry -- C$^+$ dominates in the outer PDR layers of the cloud, and
CO toward the inner shielded layers -- though the chemistry is
different.  In particular, in addition to UV radiation, cosmic rays
also contribute to the destruction of CO via the production of \hep.
As a result, for low volume density and surface density clouds, \cp
\ can dominate the carbon budget both in the outer atomic PDR, as well
as in much of the \htwo \ gas.  Toward the cloud interior, the
increased volume and surface densities within the cloud protect
against the photodissociation/ionisation of \ci \ and \chx/\ohx
\ molecules (that are principle reactants in forming CO), as well as
against the production of \hep, which is a dominant destroyer of CO
\citep{bisbas15a}.  Hence, in the innermost regions of clouds where
the surface densities are highest, the carbon is principally in
molecular CO form.

As the total column density of a cloud increases, the transition layer
between atomic and molecular (both for hydrogen and carbon) is forced
to shallower radii. This occurs because, although increasing column
density raises the SFR and thus the UV and cosmic ray intensities,
this is outweighed by the increase in cloud shielding and volume density
that accompany a rise in $\Sigma_g$. The net effect is that clouds
with high surface density and thus high star formation rate also
tend to be dominated by CO, with only a small fraction of their carbon
in the form of \cp.

\begin{figure*}
\includegraphics[scale=1]{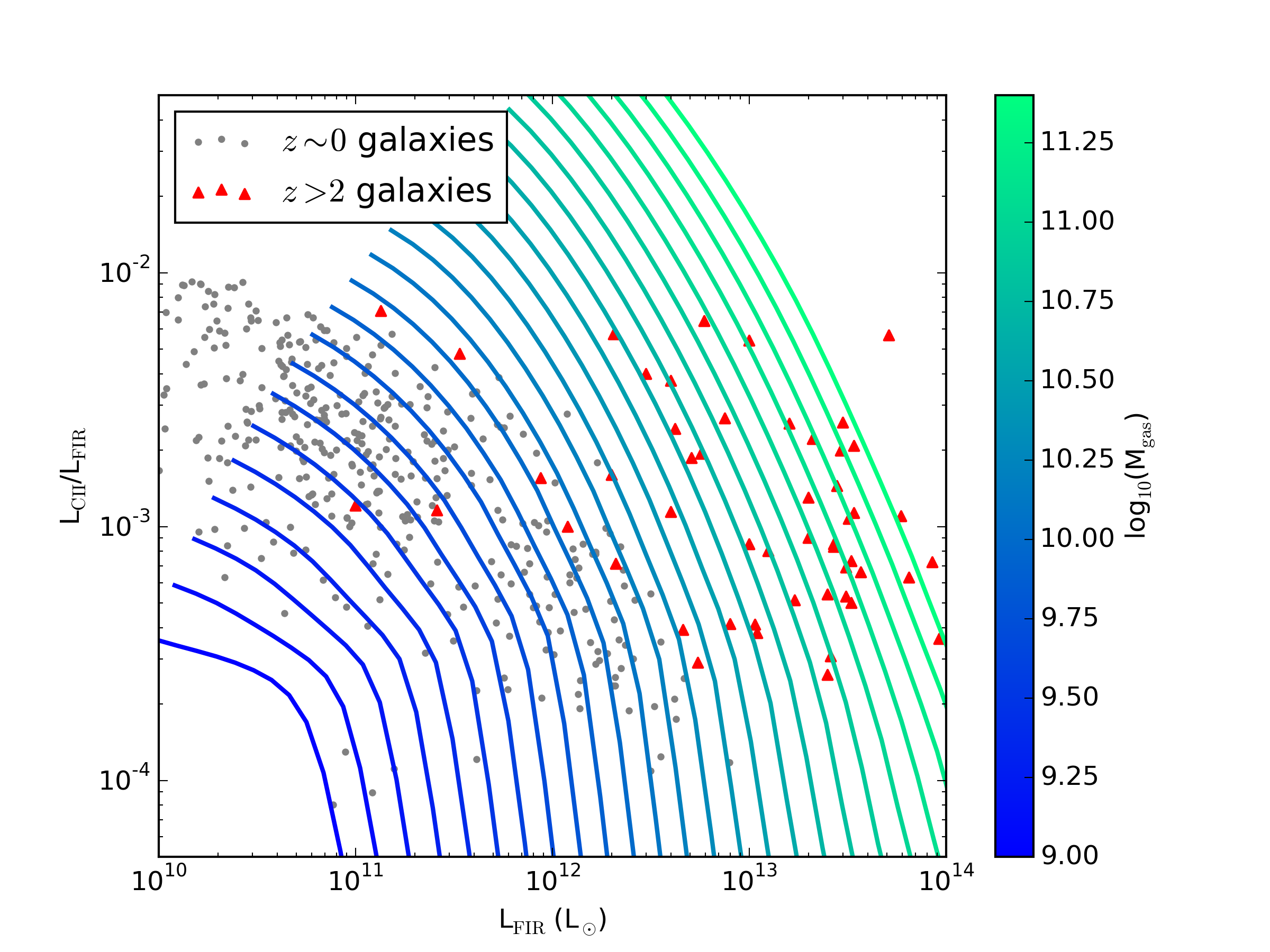}
\caption{Theoretical \cii \ luminosities (normalised by the FIR
  luminosity) as a function of FIR luminosity.  The model tracks show
  predictions for galaxy gas masses log$_{10}$(M$_{\rm gas}$) =
  [9,11.3].  The grey points show local \zsim 0 data, and the red
  triangles show high-$z$ data.  The lowest mass model track
  corresponds to the left-most one.  The \cii-FIR deficit in galaxies
  owes principally to a decrease in PDR mass in galaxies with
  increasing infrared luminosity.  High-redshift galaxies are observed
  to be systematically at a higher infrared luminosity at a given
  \cii/FIR luminosity ratio as compared to low-$z$ galaxies. In our
  model, this arises because galaxies at high-$z$ have systematically
  larger gas masses.  At a fixed star formation rate, an increased gas
  mass means lower cloud surface densities on average, which results
  in higher \cii \ luminosities.  High-$z$ detections are from
  \citet{cox11a,debreuck11a,george13a,graciacarpio11a,ivison10a,maiolino05a,rawle13a,stacey10a,swinbank12a,valtchanov11a,venemans12a,wagg12a,wang13a,willott13a,diazsantos15a,schaerer15a,gullberg15a}.  \label{figure:cii_fir_masses}}
\end{figure*}

\subsection{The \cii-FIR Relation}
\label{section:cii_fir_deficit}

We are now in a position to compare our full model to the observed
\cii-FIR relation. We do so in \autoref{figure:cii_fir_masses}, using
a large range of galaxy gas masses in our model, chosen to be
representative of the typical gas mass range of both local galaxies
and high-\z \ galaxies \citep{saintonge11a,bothwell13a,casey14a}.  We
compare these model tracks to observational data from both
low-redshift galaxies (grey points), as well as $z>2$ galaxies (red
triangles).  We discuss the local deficit relation here, and defer
discussion of the high-$z$ data to \S~\ref{section:high_redshift}. The
local data is comprised of a compilation by \citet{brauher08a}, as
well as more recent \zsim 0 data taken by \citet{diazsantos13a} and
\citet{farrah13a}.  Two trends are immediately evident from
\autoref{figure:cii_fir_masses}: (1) at fixed FIR luminosity, the
\cii/FIR ratio increases at larger galaxy gas mass, and (2) at
increasing FIR luminosity, the \cii/FIR ratio decreases.  We discuss
these trends in turn.

The trend with galaxy gas mass is straight forward to understand. The
total star formation rate is an increasing function of both gas mass
and gas surface density. Thus an increase in gas mass at fixed SFR
corresponds to a decrease in $\Sigma_g$. Because $\Sigma_g$ is a
primary variable controlling the chemical balance between \cp\ and CO,
this in turn leads to an increase in the \cp\ abundance.  The net
effect is that, at fixed SFR (and hence FIR luminosity), higher gas
mass galaxies have stronger \cii\ emission.

The second broad trend, the decrease in the \cii-FIR ratio with
increasing FIR luminosity, is the so-called \cii-FIR deficit.  The
origin of this is evident from examining the trends in both the gas
physical properties and chemical abundances in the ISM calculated thus
far (e.g. \autoref{figure:abundances}). \cp\ dominates the
weakly-shielded PDR layers of giant clouds in the ISM, while CO
principally resides in the well-shielded cloud interiors. Thus an
increase in $\Sigma_{\rm g}$ drives a decrease in the amount of
\cp\ and an increase in the amount of CO. At the same time, an
increase in $\Sigma_{\rm g}$ drives an increase in SFR and thus in FIR
luminosity. Thus an increase in $\Sigma_{\rm g}$ leads to a sharp fall
in the ratio of \cii/FIR. In an actual sample of galaxies the ratio of
\cii~to FIR falls only shallowly with FIR, however, because the
dependence on $\Sigma_{\rm g}$ is partly offset by the dependence on
gas mass. That is, galaxies with higher FIR luminosities tend to have
both higher gas surface densities and higher gas masses than galaxies
with lower FIR luminosities.  The former drives the \cii~luminosity
down and the latter drives it up, but the surface density dependence
is stronger (due to the exponential nature of FUV attenuation),
leading to an overall net decrease in \cii~emission with FIR
luminosity in the observed $z\sim 0$ sample.

\section{Discussion}
\label{section:discussion}

\subsection{Calorimetry of Giant Clouds}
\label{section:calorimetry}

  Nominally, \cii \ line cooling is one of the principle
coolants of the netural interstellar medium.  Because the \cii \ line
emission does not increase in proportion to the star formation rate,
it is interesting to consider where the cooling occurs in place of the \cii
\ line.

\begin{figure}
\hspace{-0.5cm}
\includegraphics[scale=0.45]{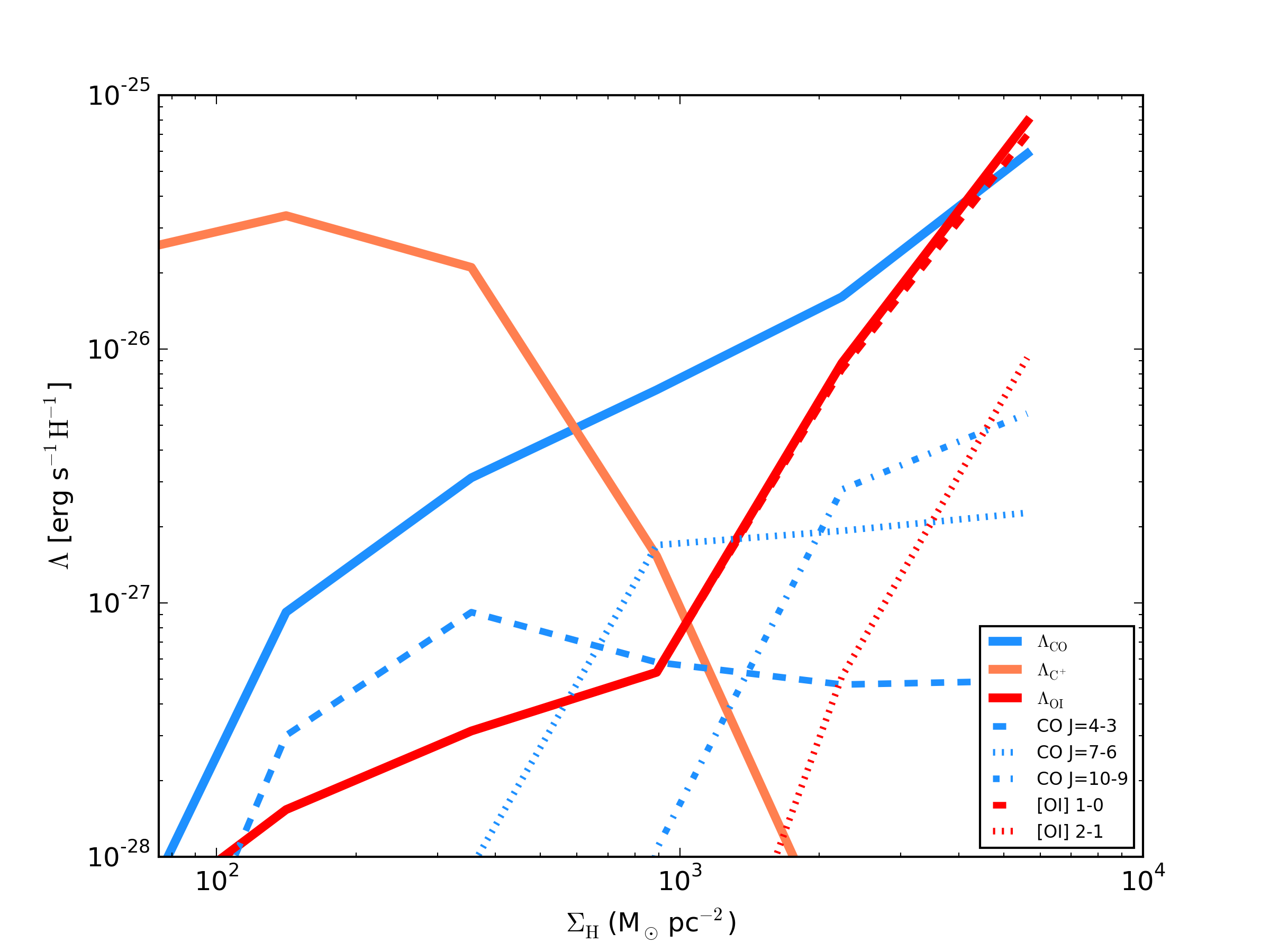}
\caption{Cooling rates of individual lines as a function of cloud
  surface density for a model galaxy of mass $M_{\rm gas} = 10^9$
  \msunend.  As the \cii \ luminosities decrease with increasing gas
  surface density, the principle cooling transitions to CO and \oi
  \ emission lines.  At high surface densities, the CO cooling is
  dominated by high-J rotational transitions.  Note, because we only
  show a subset of the CO lines for clarity, the sum of the cooling
  rates of the shown CO lines won't add up to the total CO cooling
  rate shown.
  \label{figure:calorimetry}}
\end{figure}

  In \autoref{figure:calorimetry}, we revisit the cooling rates
  originally presented in \autoref{figure:physical_properties_sd}.
  For clarity, we omit the heating rates, but additionally show the
  cooling rates from a subset of the individual CO and \oi \ emission
  lines.  As the cloud surface densities increase, the cooling rate of
  \cii \ decreases dramatically.  This owes principally to the
  plummeting \cp \ abundances.  At the same time, the dominant line
  cooling transitions to \oi \ and CO.  The increase in CO line
  cooling in part due to the rapid increase in CO abundance as the
  increased cloud surface density protects the molecule from
  photodissociation, and increased volume density combats dissociation
  via \hep.  The CO cooling is dominated by mid to high-J CO emission
  lines, with the power shifting to higher rotational transitions at
  higher gas surface densities \citep[e.g.,][]{narayanan14a}.

\begin{figure}
\hspace{-0.5cm}
\includegraphics[scale=0.5]{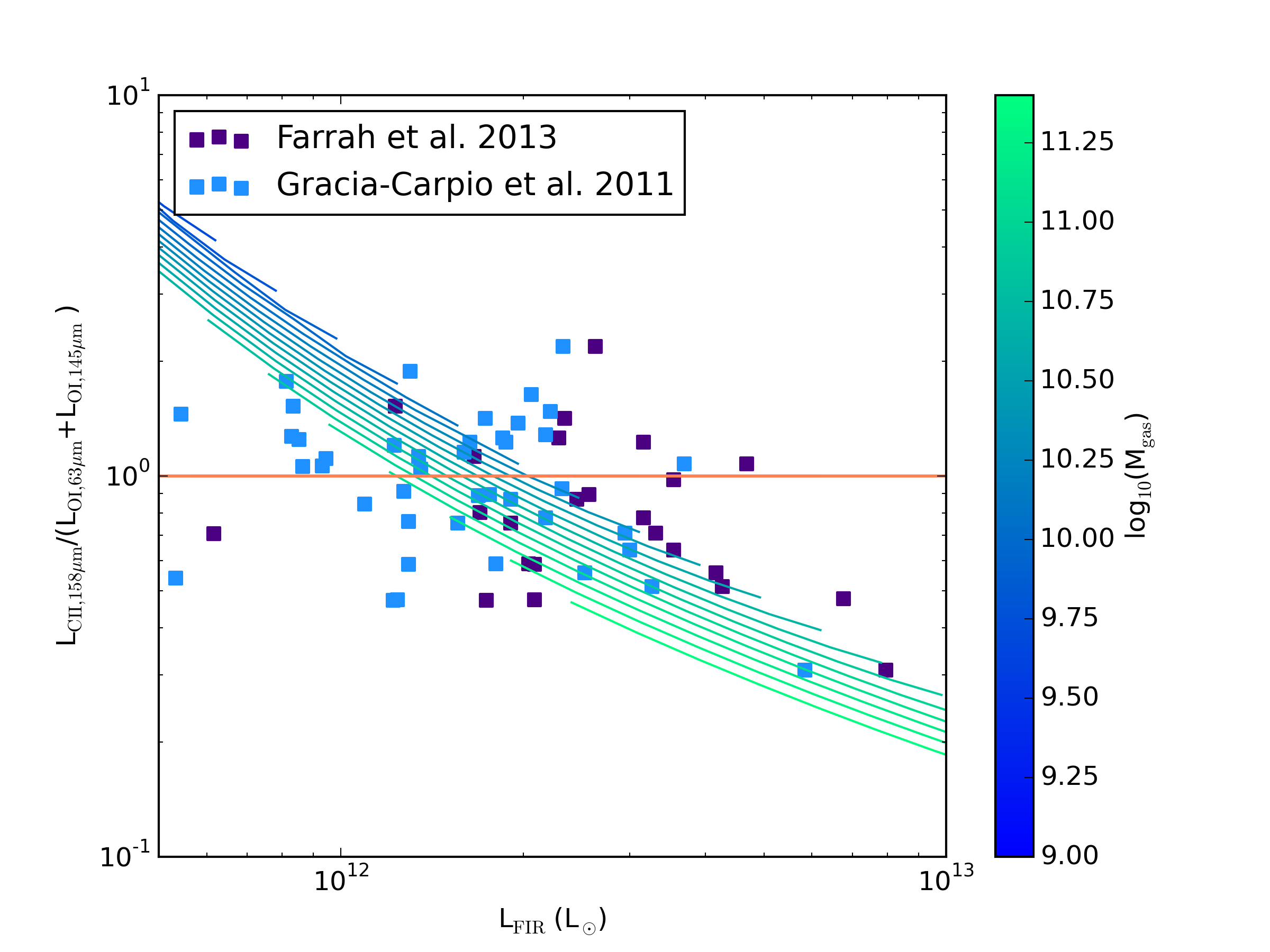}
\caption{\cii/\oi \ cooling rates for model compared to data from
  \citet{graciacarpio11a} and \citet{farrah13a}.  The \oi \ cooling
  rates are a sum of the $63 \mu$m and $145 \mu$m emission lines.
  At high gas surface density (or galaxy infrared luminosity), the
  dominant cooling in the model transitions from \cii \ line emission
  to \oi \ and CO line emission.  The orange line shows where the
  cooling power transitions from \cii \ to \oi. \label{figure:oi_fir}}
\end{figure}

Alongside CO, line emission from \oi \ is an important contributor at
high cloud surface densities.  To see why, consider again the radial
abundances within a sample cloud presented in
\autoref{figure:abundances}.  Here, we now highlight the \oi
\ abundance gradients.  While the fractional \oiatom \ abundance
decreases modestly with increasing gas surface density owing to
increased molecule production (mostly CO and \ohx), \oiatom \ remains
relatively pervasive in both atomic and molecular gas.  This is to be
contrasted with ionised \cp, which tends to reside principally in the
atomic PDRs of clouds, and sometimes the outer layer of the \htwo
\ core. So, while the mass fraction of \htwo \ to \hi \ gas increases
with increasing surface density clouds, \oi \ remains an efficient
coolant.  The bulk of the cooling occurs via the \oi \ 1-0 transition,
though emission from the 2-1 \oi \ line can be non negligible at the
highest gas surface densities.

The contribution of \oi \ line emission to the total cooling
constitutes a prediction of our model.  We quantify this in
\autoref{figure:oi_fir}, where we show the \cii/\oi \ luminosity ratio
for our model compared to observations by \citet{graciacarpio11a} and
\citet{farrah13a} of $z \sim 0.1-0.3$ galaxies.  The \oi
\ observations are of the sum of the 63 $\mu$m and 145 $\mu$m line
luminosities. At high cloud surface densities, the model luminosity
ratio between \cii \ and \oi \ decreases.  Whether such a trend exists
in the data is unclear, and requires additional data (especially at
the high-luminosity end).

\subsection{Application to High-Redshift Galaxies}
\label{section:high_redshift}

Recent years have seen a large increase in the number of \cii
\ detections from heavily star-forming galaxies at $z \ga 2$
\citep[e.g.,][]{haileydunsheath08a,stacey10a,brisbin15a,gullberg15a,schaerer15a},
extending to the epoch of reionisation
\citep[e.g.,][]{riechers13a,rawle14a,wang13a,capak15a}.  Returning to
\autoref{figure:cii_fir_masses}, we now highlight the high-redshift
compilation denoted by the red triangles. The compilation is
principally culled from the \citet{casey14a} review article, with
some more recent detections. We exclude data that have upper limits on
either \cii \ or FIR emission.  It is clear that the high-\z \ data
are offset from low-\z \ galaxies such that at a fixed \cii/FIR
luminosity ratio, \zsim 2-6 galaxies have a larger infrared
luminosity.  One possible interpretation of the high-$z$ data is that
the high-$z$ galaxies exhibit a \cii-FIR deficit akin to that observed
in local galaxies, but shifted to higher luminosities.

We now highlight the model tracks overlaid for galaxies of mass
between $M_{\rm gas} = 10^{10}-10^{11.3}$ \msunend.  These gas masses
are chosen based on the range of \htwo \ gas masses constrained for a
sample of high-$z$ submillimetre galaxies by \citep{bothwell13a}.  As
is evident from \autoref{figure:cii_fir_masses}, the model tracks for
galaxies with large gas mass show good correspondence with the
observed high-$z$ data points.  This suggests that the ultimate reason
for the offset in infrared luminosity for the \cii-FIR deficit of
high-$z$ galaxies is their large gas masses. High infrared
luminosity-galaxies at $z\sim 0$ typically have high star formation
rates due to large values of $\Sigma_g$, but those at high-$z$ have
high star formation rates due to large gas masses instead.

Our interpretation for the offset in infrared luminosity in the
high-$z$ \cii-FIR deficit is consistent with a growing body of
evidence that, at a fixed stellar mass, galaxies at high-redshift have
higher star formation rates and gas masses than those at $\zsim 0$
\citep{dave10a,geach11a,rodighiero11a,madau14a,narayanan12c,narayanan15a},
and that it is these elevated gas masses that are driving the extreme
star formation rates, rather than a short-lived starburst event.  As
an example, the most infrared-luminous galaxies at \zsim 0 have small
emitting areas ($\sim 1 $ kpc), and large measured gas surface
densities, up to $\sim 10^3$ \msun pc$^{-2}$ averaged over the emitting
area.  In contrast, galaxies of comparable luminosity at high-$z$ have
a diverse range of sizes, with some gas spatial extents observed $\sim
20$ kpc \citep{casey14a,ivison11a,simpson15a}. Indeed, cosmological
zoom simulations by \citet{narayanan15a} have shown that the extreme
star formation rates of the most infrared-luminous galaxies $L_{\rm
  IR} \sim 10^{13}$ \lsun at \zsim 2 can be driven principally by
significant reservoirs of extended gas at a moderate surface density.
Similarly, at lower luminosities at $\zsim 2$, \citet{elbaz11a} find
that the cold dust SEDs of main sequence galaxies are consistent with
more extended star formation at lower surface densities than their
low-$z$ counterparts.  Our model suggests that the offset in the
\cii-FIR relation between $z\sim 0$ and $z\sim 2$ galaxies can
ultimately be traced to the same phenomenon.

\subsection{Relationship to other Theoretical Models}
\label{section:theory}
There has been significant attention paid to modeling \cii \ emission
from galaxies over the past 5-10 years.  The methods are broad, and
range from numerical models of clouds, as in this paper, to
semi-analytic dark matter only simulations to full cosmological
hydrodynamic calculations.  Generally, models have fallen into three
camps (with some overlap): (1) Models that study the \cii-SFR
relationship in low luminosity galaxies at a given redshift
\citep[e.g.,][]{olsen15a}; (2) models predicting \cii \ emission from
epoch of reionisation galaxies
\citep{nagamine06a,munoz13a,munoz13b,vallini13a,vallini15a,pallottini15a};
and (3) models that aim to understand the \cii-FIR deficit in luminous
galaxies.  We focus on comparing to other theoretical models in this
last category as they pertain most directly to the presented work
here.

\citet{abel09a} use {\sc cloudy} H~\textsc{ii} region models
\citep{ferland13a} to explore the idea, first posited by \citet{luhman03a},
that \cii~emission is suppressed in high-SFR galaxies because H~\textsc{ii}
regions become dust-bounded rather than ionisation bounded. In their
model, this reduces the \cii~luminosity because ionising photons absorbed
by dust are not available to ionise carbon. This model implicitly assumes that
\cii~emission arises predominantly from ionised regions and the PDRs that
surround them, because it does not consider \cii~emission coming from
phases of the ISM that are not associated with the immediate environments
of stellar clusters young enough to drive H~\textsc{ii} regions. However,
observations suggest that H~\textsc{ii} regions and PDRs account
for only $\sim 50\%$ of the \cii~emission in the Milky Way, with that fraction
declining at higher SFRs \citep[their Figure 6 in particular]{pineda14a}. Thus
it is unclear if a model based around H~\textsc{ii} regions is capable of
explaining the galaxy-averaged \cii-FIR relation, which is dominated by
other ISM phases.

\citet{bisbas15a} developed cloud models with similar underlying
methods to those presented here, based on the  {\sc 3d-pdr} code.
They use their models to investigate the chemistry of CO, and find, as
we do, that at high star formation rates \hep\ destruction of CO becomes
an important process in determining the overall carbon chemical balance
in a galaxy. However, they do not consider \cii\ emission or its relationship
with star formation rates. Similarly,  \citet{popping13b} developed
semi-analytic galaxy formation models
(SAMs) coupled with PDR modeling to model CO, \ci \ and \cii
\ emission from model galaxies.  The models provide a reasonable match
to the observed \zsim 0 \cii-FIR deficit, but the authors do not discuss
the physical origin of the effect, nor its redshift dependence.

Finally, \citet{munoz15a} posited an analytic model in which \cii \ line
saturation may drive the observed \cii/FIR luminosity deficit.  At
very high temperatures ($T_{\rm gas} \gg T_{\rm [CII]}$), 
the line luminosity no longer increases with temperature, and they
suggest that this phenomenon is why \cii~stops increasing with SFR
at high SFRs. However, our models suggest that the gas temperatures
typically do not reach such high values, at least up to
 $L_{\rm FIR} \sim 10^{13}$.  For example, turning to
\autoref{figure:physical_properties_sd}, we find
maximum\footnote{This number of course increases with increasing
  galaxy gas mass.  Functionally, we find a maximum temperature of $T
  \sim 100 K$ for our most massive galaxies.} gas temperatures of
$\sim 50$ K at luminosities of $L_{\rm FIR} = 10^{13}$ \lsunend.
Further bolstering this claim, \citet{narayanan11b} ran idealised
hydrodynamic galaxy merger simulations with thermal balance models
similar to the ones presented in this paper, though with the increased
sophistication of directly modeling the dust temperature via 3D dust
radiative transfer calculations.  \citet{narayanan11b} found that even
for clouds in merger-driven starbursts that exceed $\Sigma_{\rm H2} >
10^4$ \msun pc$^{-2}$, the gas kinetic temperature remained $T_{\rm
  kin} \sim 10^2$ K.  Thus, within the context of the thermal balance
equations presented in this work, we conclude that line saturation is
not likely to dominate the \cii/FIR luminosity deficit at high
infrared luminosities.

\subsection{Relationship to Observations}

There are a few observational works in recent years that discuss
particular aspects of the \cii/FIR luminosity deficit with FIR
luminosity that are worth discussing within the context of the model
that we present in this paper.  In early work, \citet{stacey91a}
utilised PDR models to interpret their KAO observations of a sample of
nearby gas-rich galaxies.  These authours found that single-component
PDR models did not fit their observed \cii/FIR luminosity ratios, and
posited a two-component model in which the bulk of the FIR emission
arises from a confined region exposed to a high intensity UV field,
while the \cii \ line emission mostly comes from a more diffuse
extended component exposed to a lower radiation field. This is 
compatible with the model we present here.  While in our model we
assume for simplicity that a given galaxy is comprised of a single
population of clouds, in reality a galaxy will have a distribution of
cloud surface densities, likely with the high $\Sigma_{\rm g}$ clouds
toward the central regions associated with more active star formation.
In this scenario, the FIR luminosity would predominantly arise from
the high $\Sigma_{\rm g}$ clouds, while the lower $\Sigma_{\rm g}$
clouds would produce more \cii \ line emission, consistent with the
model developed by \citet{stacey91a}.

\citet{diazsantos13a} presented
Herschel observations of 241 nearby galaxies, and confirmed the
\cii-FIR deficit in their relatively large sample.  They find that the
luminous infrared galaxies (LIRGs) with higher infrared luminosity
surface density tend to have lower \cii/FIR luminosity ratios compared
to more extended systems.  Within the context of our model, we
interpret this result as higher infrared luminosity density galaxies
corresponding to higher gas surface density galaxies
\citep{kennicutt12a}.  Galaxies with higher gas surface densities will
have lower atomic PDR masses with respect to their \htwo \ molecular
masses, and therefore lower \cii \ luminosities.

\subsection{Uncertainties in the Model}
\label{section:uncertainties}

\begin{figure}
\hspace{-0.5cm}
\includegraphics[scale=0.5]{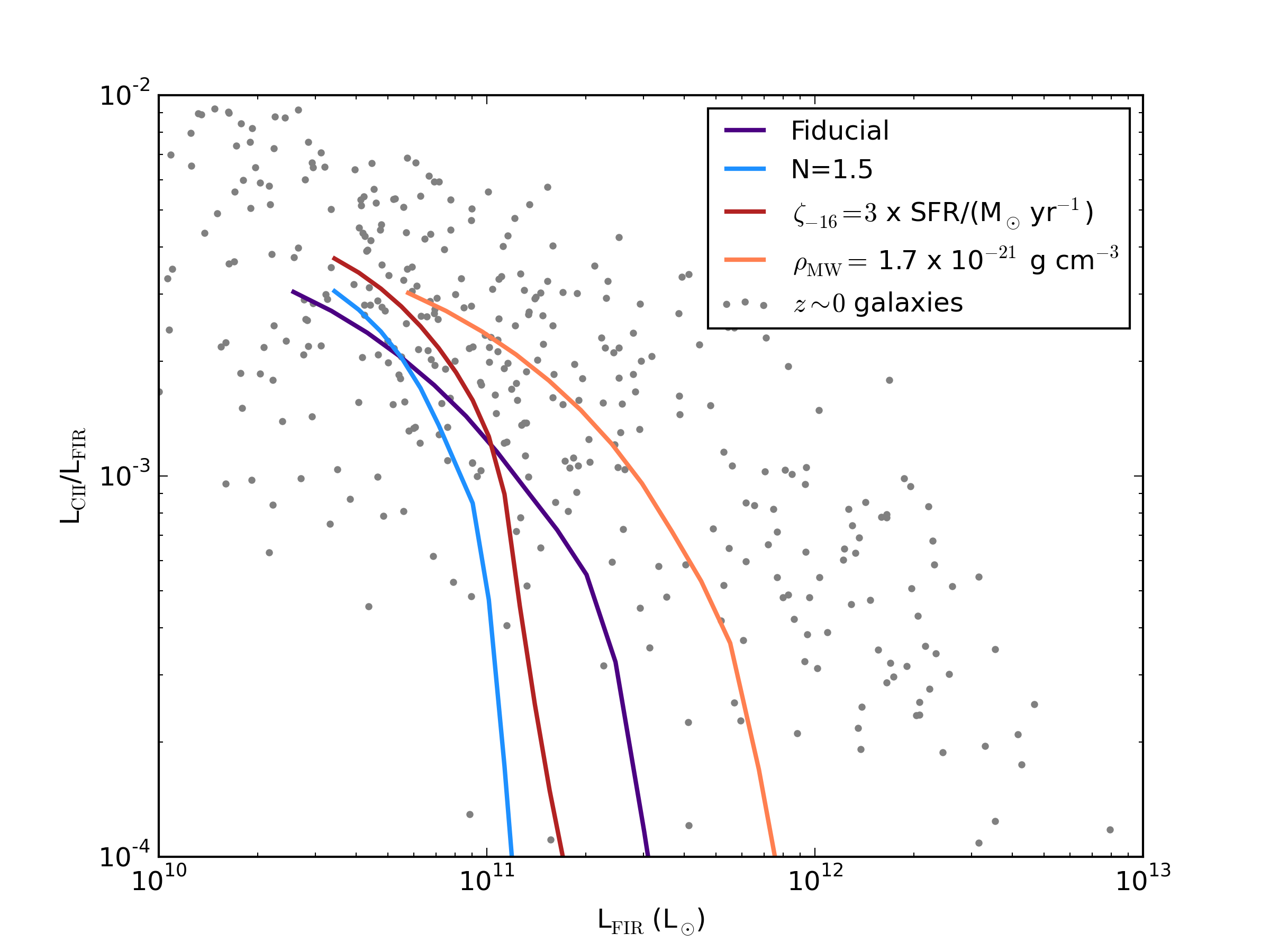}
\caption{\cii/FIR luminosity ratio versus FIR luminosity for our
  fiducial model (thick blue), and three varying parameters in our
  cloud models.  The example shown is for a $M_{\rm gas} = 3 \times
  10^{9} \msun$ galaxy, the average mass (in log-space) of our
  low-redshift galaxy mass range.  See
  Table~\ref{table:free_parameters} for variable definitions.
  \label{figure:parameter_survey}}
\end{figure}

While we aim to provide a relatively minimalist analytic model for the
structure of giant clouds in galaxies in \autoref{section:model},
there are a number of free parameters that we had to choose that can
impact the overall normalisation of the \cii/FIR ratio in our model
galaxies.  In this section, we explore the impact of these parameters
on our results.  In \autoref{figure:parameter_survey}, we summarise
the impact of varying the star formation law index ($N$), the density
normalisation ($\rho_{\rm MW}$), and the cosmic ray ionisation rate
($\zeta_{-16}$).

All of these values are uncertain.  Star formation indices ranging
from sublinear to quadratic have been reported in the literature
\citep{kennicutt12a}.  Decreasing the star formation law index
decreases the average density of our model clouds which has a two-fold
effect.  First, it decreases the total star formation rate via longer
cloud free fall times.  Second, \cii \ is more easily formed owing to
decreased recombination and molecular formation rates.  Increasing the
cloud density normalisation ($\rho_{\rm MW}$) has the opposite effect.

Similarly, the cosmic ray ionisation rate in galaxies is uncertain.
Observations of H$_3^+$ within the Galaxy suggest rates that range
from $\zeta_{-16} \approx 0.1-3$
\citep{mcall99a,indriolo07a,neufeld10a,wolfire10a}.  For lack of a
better constraint, we have chosen to employ a relatively conservative
value ($\zeta_{-16} = 0.1 \times$ SFR/(M$_\odot$ yr$^{-1}$), noting
that increased cosmic ray ionisation rates tend to result in more \cii
\ flux.  This is shown explicitly in
\autoref{figure:parameter_survey}, where we test a model with
$\zeta_{-16} = 3 \times $SFR/(M$_\odot$ yr$^{-1}$).

For all of these tested variations in our parameter choice survey, the
overall trend is similar: the \cii/FIR luminosity ratio decreases with
increasing galaxy star formation rate.  This is because the net
decrease in the \cii/FIR luminosity ratio owes to decreasing relative
PDR masses in increasingly luminous galaxies in our model.  The
normalisation and exact behaviour of course depends on uncertain
parameters, but the broad trends are robust.

 Finally, we note note that the linear scaling of
the cosmic ray ionisation rate with star formation rate is rooted in
the tentative observational evidence for such a scaling by
\citet{abdo10b}.  However, it is plausible that the cosmic ray
ionisation rate scales with some other parameter, such as star
formation rate density or surface density.  We defer a more thorough
investigation into the chemical consequences of varying ionisation
rate scalings to a forthcoming study.

\section{Summary}
\label{section:summary}

We have developed models for the physical structure of star-forming
giant clouds in galaxies, in which clouds consist of both atomic and
molecular hydrogen.  We coupled these cloud models with chemical
equilibrium networks and radiative transfer models in order to derive
the radial chemical and thermal properties, and utilised these models
to investigate the origin of the \cii/FIR luminosity deficit in
luminous galaxies.  Our main results follow:
\begin{enumerate}
\item The carbon-based chemistry in clouds can be reduced to a
  competition between volumetric density and surface density
  protecting the formation of molecules against cosmic-ray induced
  ionisations, and UV photodissociations/ionisations.

  \item At solar metallicities, the molecular \htwo \ core in clouds
    tends to be well-shielded from UV radiation, and dense enough to
    combat CO destruction by cosmic rays (via an intermediary \hep
    \ reaction) as well as ultraviolet radiation. As a result, the
    carbon in molecular gas is principally in the form of CO.  This
    said, the lower surface density surfaces of the central molecular
    core can contain significant amounts of \cp.

    \item The opposite is true for atomic PDRs.  The carbon in the PDR
      shell in clouds is easily exposed to the radiation and cosmic
      ray field, resulting in the bulk of carbon being in ionised \cp
      \ form.

    \item At increasing galaxy star formation rate (and infrared
      luminosity), the typical cloud surface density ($\Sigma_{\rm
        g}$) rises; with increasing $\Sigma_{\rm g}$, the molecular
      fraction (\htwo/\hi) rises, and the relative \cii \ luminosity
      decreases owing to a shrinking of the size of the \cii
      \ emitting region.  These reduced PDR masses in increasingly
      luminous galaxies are the fundamental origin of the \cii/FIR
      luminosity deficit with FIR luminosity in galaxies in our model.

    \item As the \cii \ luminosity decreases in high surface density
      clouds, the major coolants of the ISM transition to high-J CO
      emission lines and \oi.

    \item At a fixed star formation rate, galaxies of increasing gas
      mass have lower cloud surface densities, and hence larger \cii
      \ luminosities.  Because of this, galaxies at high-redshift,
      which have larger gas fractions than their low-$z$ counterparts,
      lie offset from low-$z$ galaxies in \cii-FIR space.  Our model
      provides a natural explanation for the offset \cii-FIR deficit
      for high-$z$ galaxies.

\end{enumerate}

\section*{Acknowledegements}
The authours thank Thomas Bisbas, Tanio Diaz-Santos, Duncan Farrah,
and Munan Gong, Javier Gracia-Carpio and Eve Ostriker for providing
both simulation and observational data that aided us when developing
these models.  D.N. thanks Aaron Evans, Thomas Greve, Rob Kennicutt,
Dominik Riechers, J.D. Smith, Gordon Stacey, Dan Stark, and Karen
Olsen for helpful conversations. Partial support for DN was provided
by NSF AST-1009452, AST-1442650, NASA HST AR-13906.001 from the Space
Telescope Science Institute, which is operated by the Association of
University for Research in Astronomy, Incorporated, under NASA
Contract NAS5-26555, and a Cottrell College Science Award, awarded by
the Research Corporation for Science Advancement. MRK acknowledges
support from US NSF grant AST-1405962 and Australian Research Council
grant DP160100695.

\begin{appendix}
\section{Resolution Tests}
\label{section:resolution}
\begin{figure}
  \includegraphics[scale=0.5]{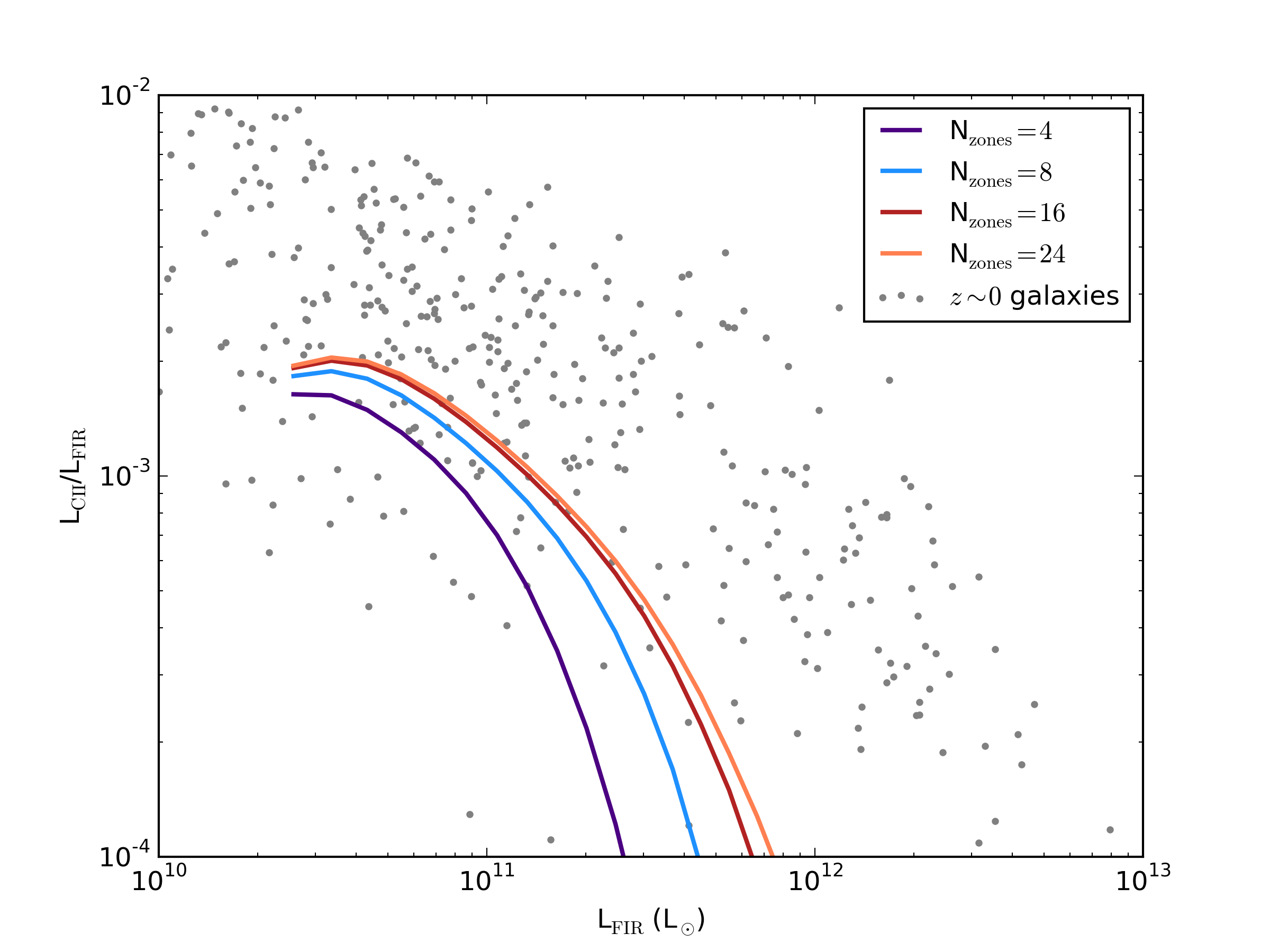}
  \caption{Convergence test for number of radial zones in model
    clouds.  Test shows model \cii-FIR deficit for a galaxy of mass
    $M_{\rm gas} = 3 \times 10^9$ \msun, the average mass (in
    log-space) of our low-redshift galaxy mass
    range. \label{figure:convergence}}
\end{figure}

Our fiducial model clouds are subdivided into 16 radial zones. In
Figure~\ref{figure:convergence}, we test the convergence properties of
this model by varying the number of zones from 4-24 for a model galaxy
of log($M_{\rm gas}$) = $9.5 $ \msun.  While differences exist at
the lowest \cii \ luminosities between $N_{\rm zone} = 16$ and $N_{\rm
  zone} = 24$, they are minor.

\end{appendix}

\bibliographystyle{mnras}
\bibliography{cii_refs}

\end{document}